\documentclass[12pt]{iopart}

\usepackage{graphicx}

\usepackage[]{youngtab}

\def\al{\alpha}
\def\be{\beta}
\def\ga{\gamma}
\def\de{\delta}
\def\ep{\epsilon}
\def\ze{\zeta}
\def\et{\eta}\def\la{\lambda}

\def\th{\theta}

\def\ka{\kappa}
\def\rh{\rho}

\def\si{\sigma}

\def\ph{\phi}

\def\ch{\chi}
\def\ps{\psi}
\def\om{\omega}

\def\Ph{\Phi}

\def\Om{\Omega}

\def\cG{{\cal G}}
\def\cL{{\cal L}}

\def\cD{{\cal D}}

\def\mn{{\mu\nu}}
\def\ab{{\al\be}}

\def\abgd{{\al\be\ga\de}}

\def\koo{{k_{00}}}
\def\kjj{{k_{ll}}}
\def\kfo{ k^{(4)} }
\def\kfi{ k^{(5)} }
\def\ksi{ k^{(6)} }

\def\hb{\bar{h}}

\def\tG{{\tilde G}}

\def\prt{\partial}
\def\pt#1{\phantom{#1}}

\newcommand{\beq}{\begin{equation}}
\newcommand{\eeq}{\end{equation}}
\newcommand{\bea}{\begin{eqnarray}}
\newcommand{\eea}{\end{eqnarray}}
\newcommand{\rf}[1]{(\ref{#1})}

\def\fr#1#2{{{#1} \over {#2}}}
\def\frac#1#2{{\textstyle{{#1}\over {#2}}}}

\begin{document}

\title[]{Short-range forces due to Lorentz-symmetry violation}

\author{Quentin G.\ Bailey$^{1}$, Jennifer L.\ James$^{2}$, Janessa R.\ Slone$^{1}$, and Kellie O'Neal-Ault$^{1}$}

\ead{baileyq@erau.edu}

\address{$^{1}$Embry-Riddle Aeronautical University, 3700 Willow Creek Road, Prescott, AZ, 86301, USA}

\address{$^{2}$Vanderbilt University, 2201 West End Avenue, Nashville, TN, 37235 USA}

\begin{abstract}
Complementing previous theoretical and experimental work, 
we explore new types of short-range modifications to Newtonian gravity arising from spacetime-symmetry breaking. 
The first non-perturbative, i.e., 
to all orders in coefficients for Lorentz-symmetry breaking, 
are constructed in the Newtonian limit.
We make use of the generic symmetry-breaking terms 
modifying the gravity sector
and examine the isotropic coefficient limit.
The results show new kinds of force law corrections, 
going beyond the standard Yukawa parameterization.
Further, 
there are ranges of the values of the coefficients that could make the resulting forces large compared to the Newtonian prediction at short distances.
Experimental signals are discussed for typical test mass arrangements.
\end{abstract}

\section{Introduction}

Presently, 
the nature of gravity is still largely 
unknown on length scales less than micrometers.
In fact, 
new types of forces many times stronger than the Newtonian gravitational force could exist on short length scales 
and still be consistent with current experimental limits 
\cite{Lee20}.
Suggestions for hypothetical new forces that could modify gravity at short ranges abound in the literature \cite{Fujii:1971vv,Donoghue:1995cz,Arkani-Hamed:1998jmv,Fischbach:1999bc,Krause:1999ry,Murata:2014nra,ahn03}.
In particular, 
miniscule but potentially detectable 
violations of fundamental symmetries underlying General Relativity (GR) can arise in a plethora of ways \cite{ksstring89,gp99,chkl01,Mattingly:2005re,Tasson:2014dfa,Addazi:2021xuf,mariz22}. 
The breaking of local Lorentz symmetry, 
for instance, 
can modify gravity on short ranges while being consistent with longer range measurements \cite{bkx15,km17}.

To categorize the phenomenology of spacetime symmetry breaking one needs a comprehensive test framework.
Effective field theory (EFT) is a widely used tool
for describing potentially detectable new physics
\cite{Weinberg:2009bg}.
EFT descriptions of spacetime-symmetry breaking, 
including local Lorentz symmetry breaking,
are based on including the action of GR 
and a standard matter sector action \cite{kp95}.
To these basic pieces, 
are added a series of symmetry breaking terms that can be organized
by number of derivatives, curvature, 
mass dimensions, and so on \cite{ck97,ck98,k04}.
This approach has the advantage that one can in principle
calculate the effect on some observable due to some symmetry breaking terms, 
which can then be compared with entirely different observables in different scenarios, 
for measurements of the same coefficients controlling the size of the effects.
Other formalisms for testing symmetries in gravity are parametrized directly from the form of a GR observable \cite{Cornish:2011ys,myw12,Will:2014kxa}, 
or are based on specific models of alternatives to GR
\cite{Jacobson:2000xp,Alexander:2009tp,Berti:2015itd,Okounkova:2021xjv}. 

We will consider in this work modifications 
to the gravity sector that, 
contrary to standard GR, 
break local Lorentz symmetry 
and diffeomorphism symmetry explicitly or spontaneously.
These spacetime symmetries can be thought of as gauge symmetries
for gravity, 
and thus GR is a gauge theory of gravity with 
local Lorentz and diffeomorphism symmetries as the gauge symmetries, 
analogous to Standard Model physics based on gauge groups \cite{Hehl:1976kj}.
The subtle issue of the role of broken spacetime symmetries
in the context of curved spacetime, 
particularly when assuming asymptotically flat scenarios or not, 
has been discussed at length elsewhere \cite{k04,bluhm15,kl21}.
While we do not fully discuss these concepts 
and subtleties here, 
we shall refer to conventions and categories of
transformations in these references as needed.

In the EFT approach taken here,
we highlight comparison of short-range (SR) gravity tests
with gravitational wave (GW) observations,
thus comparing two tests ``across the universe" 
for measuring the same quantities describing 
spacetime-symmetry breaking for gravity. 
In fact, 
we show certain rotational scalar coefficients 
that can be measured in GW tests can also be probed in SR tests.
Further, 
there are some coefficients that 
cannot be completely disentangled with GW tests alone, 
but using also SR gravity tests could accomplish this.

In references \cite{bkx15} and \cite{km17} solutions 
for short-range gravity tests were found, 
but these used an approximation of leading order in the coefficients.
We show here that exact, non-perturbative, 
solutions can reveal where other combinations of coefficients, 
not yet disentangled, 
can show up in experiment.
As we are concerned in this paper with 
modifications to gravity that do not break 
the Weak-Equivalence Principle, 
we do not discuss WEP violations here.
The connection between Lorentz violation and WEP
has been discussed at length elsewhere \cite{Fischbach:1985wq,kt09,kt11,Bars:2019lek}.

Since we examine non-perturbative solutions, 
the results in this work also touch on the nature of higher than second order derivatives in the action and how that might affect gravity.
For this latter topic, 
we do not attempt a comprehensive investigation of these
issues but simply note where results exhibit behavior expected of such models \cite{Podolsky:1942zz,Pais:1950za,Woodard:2015zca}, 
and how they might be consistent with perturbative approaches.

The paper is organized as follows.
In section \ref{background theory}, 
we review two commonly used EFT schemes for  the description of spacetime symmetry breaking in gravity and we discuss prior results
in short-range gravity signals for Lorentz violation.
In section \ref{isotropic Newtonian limit, exact solutions}, 
we explore non-perturbative solutions with a special case model 
to identify key features.
Following this,
we go on to solve the general EFT framework in the static, 
isotropic coefficient limit.
Features of the solutions are discussed and explained with several plots.
We discuss attempting exact solutions with
anistropic coefficients in section \ref{anisotropic exact solutions}, 
and compare to perturbative methods.
For Section \ref{experimental implications}, 
we apply the theoretical results to simulate 
the signal of the gravitational field above a flat plate of mass, 
and comment on the experimental signatures.
A summary and outlook is provided in section \ref{summary and outlook}.
Finally, 
in the appendix we include a review of relevant differential equations, 
the details of the tensor analysis for isotropic coefficients used, 
and special cases of the SR gravity solutions.
In this work, 
we assume 4 dimensional spacetime with metric signature
$-+++$ and units where $\hbar=c=1$.
Latin letters are used for 3 dimensional space, 
and Greek letters for spacetime indices.

\section{Background theory}
\label{background theory}

\subsection{Action and field equations}

One can work with an observer covariant EFT expansion
or an action designed for weak-field applications, 
the latter formulated in a quadratic action expansion.
The two approaches are overlapping descriptions of physics beyond
GR and the SM when spacetime symmetries are broken.
We display both approaches here, 
to emphasize recent points of view in the literature, 
and because we use them in this work.

It is a basic premise that in the EFT context,
a breaking of spacetime symmetries is indicated 
by the presence of a background tensor field of some kind that couples to matter or gravity or both
\cite{ksstring89,kp95,ck97}.
The details and subtleties of this premise have been discussed at length elsewhere \cite{k04,bluhm15,kl21}.
Suffice it to say here that the EFT maintains coordinate invariance of physics (observer invariance)
while the action may not be invariant under
symmetry transformations of localized field configurations (particle transformations). 
The latter violation is due to the presence of the background tensor fields, 
which remain fixed under such transformations.

The observer covariant expansion 
has a Lagrange density that takes the form of a series of terms:
\bea
     \cL &=& \frac {\sqrt{-g}}{2\ka} 
     \big( R + \kfo_\abgd R^\abgd + \kfi_{\al\be\ga\de\ka} \nabla^\ka R^{\abgd}
     \nonumber\\
     &&\pt{space}
     + \ksi_{\ka\la\mu\nu\abgd} R^{\ka\la\mu\nu} R^{\abgd}+ ... \big)
     + \cL^\prime.
\label{lag1}
\eea
In this expression, 
the determinant of the metric is $\sqrt{-g}$,
$R^\abgd$ is the Riemann curvature tensor, 
$R$ is the Ricci scalar, 
and 
$\kfo_\abgd$, 
$\kfi_{\al\be\ga\de\ka}$, 
and $\ksi_{\ka\la\mu\nu\abgd}$
are the coefficients controlling the degree of symmetry breaking \cite{k04,bkx15}.
The coupling is $\ka=8\pi G_N$, 
where $G_N$ is the gravitational constant.
The first term is the Einstein-Hilbert lagrange density, 
while the remaining terms are the symmetry-breaking terms.
Note that additional terms for the coefficients can be included in $\cL^\prime$.
For instance, 
a general expansion for such terms exists,
for the case of a two-tensor
$s_\mn \propto k^{(4)\al}_{\pt{(4)\al}\mu\al\nu}$, 
and takes the form 
\bea
\cL^\prime &=& \fr {\sqrt{-g}}{2\ka } 
\Big[ 
a_3 \frac 12 (\nabla_\mu s_{\nu\la}) (\nabla^\mu s^{\nu\la}) 
+a_4 \frac 12 (\nabla_\mu s^{\mu\la}) (\nabla_\la s^\be_{\pt{\be}\be}) 
\nonumber\\
&&
\pt{space}+...
+a_7 s_\mn s_{\ka\la} R^{\mu\ka\nu\la}
+a_8 s_\mn s^{\mu}_{\pt{\mu}\la} R^{\nu\la}
+...
\Big],
\label{dyn}
\eea
which can be viewed as terms of second order in the coefficients or as dynamical terms \cite{b21,kl21}.
Alternatives to \rf{dyn} can adopt the explicit symmetry breaking scenario, 
where the coefficients in \rf{lag1} 
are given {\it a priori}, 
this latter possibility given emphasis more recently \cite{Bonder:2020fpn,abn21,Reyes:2021cpx,Reyes:2022mvm}.

An alternative overlapping approach, 
the quadratic action approach, 
assumes an expansion around flat spacetime $\et_\mn$, 
of the standard form
\beq
g_\mn = \et_\mn + h_\mn.
\label{metric}
\eeq
We examine the quadratic action \cite{km16,km18} in the limit that maintains the usual linearized gauge invariance of GR: 
$h_{\mu\nu}\rightarrow h_{\mu\nu}-\prt_{\mu}\xi_{\nu}-\prt_{\nu}\xi_{\mu}$.
The Lagrange density for this approach takes the form
\beq
        \cL = -\frac{1}{4\ka} h^\ab G_\ab
        +\frac{1}{8\ka} h_{\mu\nu}(\hat{s}^{\mu\rh\nu\si}+\hat{q}^{\mu\rh\nu\si}
        +\hat{k}^{\mu\rh\nu\si})h_{\rh\si},
         \label{lag2}
\eeq
where $G_{\al\be}$ is the linearized Einstein tensor.
The ``hat" operators are built from background coefficients for spacetime-symmetry breaking and partial derivatives.
The three types appearing 
in \rf{lag2} are given by,
\bea
    \hat{s}^{\mu\rh\nu\si}&=&s^{(d)\mu\rh\ep_1\nu\si\ep_2...\ep_{d-2}}\prt_{\ep_1}...\prt_{\ep_{d-2}}, \nonumber\\
    \hat{q}^{\mu\rh\nu\si}&=&q^{(d)\mu\rho\ep_1\nu\ep_2\si\ep_3...\ep_{d-2}}\prt_{\ep_1}...\prt_{\ep_{d-2}}, \nonumber\\
    \hat{k}^{\mu\nu\rh\si}&=&k^{(d)\mu\ep_1\nu\ep_2\rh\ep_3\si\ep_4...\ep_{d-2}}\prt_{\ep_1}...\prt_{\ep_{d-2}}.
    \label{sqk}
\eea
While the expansions in \rf{sqk} appear similar for the three types of coefficients, 
the $s$, $q$, 
and $k$ in fact differ by symmetry and tensor properties.
The detailed tensor properties 
of these terms are described 
in the Young Tableau of Table 1 of Ref.\ \cite{km16}, 
(some samples are included in appendix \rf{spacetimeyoung}).
In particular, 
$\hat{s}^{\mu\rh\nu\si}$
is anti-symmetric in the pairs of indices $\mu\rh$ and $\nu\si$, 
while $\hat{q}^{\mu\rh\nu\si}$ 
is anti-symmetric in 
$\mu\rh$ and symmetric in $\nu\si$, 
and finally $\hat{k}^{\mu\nu\rh\si}$
is symmetric in the pairs of indices $\mu\rh$ and $\nu\si$.
In terms of discrete spacetime symmetries, 
The $\hat{s}$ operators have even CPT symmetry and mass dimension $d \geq 4$; $\hat{q}$ operators have odd CPT and mass dimension $d \geq 5$; $\hat{k}$ operators have even CPT and mass dimension $d \geq 6$.

The phenomenology of the terms in
\rf{lag1} and \rf{lag2} has been studied in a number
of works.
Observable effects in weak-field gravity
tests have been established for a subset of the possible terms \cite{bk06,Hees:2016lyw,bkx15}
and some work has been done on strong-field 
gravity regimes like cosmology \cite{Bonder:2017dpb,abn21,Reyes:2022dil,Nilsson:2022mzq}.
Effects on gravitational waves have been studied, 
showing that dispersion and birefringence
occur generically as a result of CPT and
Lorentz violation \cite{km16}.
Analysis has been performed in tests such as lunar laser ranging \cite{Bourgoin:2016ynf}, 
gravimetry \cite{Muller:2007es}, 
pulsars \cite{Shao:2014oha}, 
and using the catalog of GW events \cite{Abbott_2017,Liu:2020slm,shao20,wang21,ONeal-Ault:2021uwu}.
An exhaustive list of up to date experimental limits and papers on gravity sector coefficients can be found in 
\cite{datatables}.

On the theory side, 
explicit local Lorentz and 
diffeomorphism symmetry cases 
have been explored various contexts.
A ``3+1" formulation of the 
EFT framework 
has been explored 
in Refs.\ \cite{abn21,Reyes:2021cpx,Bonder:2022fsj}.
Extensive work has been completed mapping out the approach to explicit symmetry breaking with Finsler geometry \cite{Kostelecky:2011qz,Lammerzahl:2012kw,AlanKostelecky:2012yjr,Schreck:2014hga}.
Other work includes much attention 
to vector and tensor models of spontaneous symmetry breaking \cite{Jacobson:2000xp,bk05,bk08,Horava:2009uw,Alexander:2009tp,Kostelecky:2009zr,abk10}
and how these models can be matched to the EFT expansion above \cite{seifert09,Bluhm:2019ato,abn21,Reyes:2022mvm}.
More recently,
black hole solutions have been studied \cite{Eling:2006ec,casana18,Xu:2022frb}.
Also, 
the systematic construction of dynamical terms for the spontaneous symmetry breaking scenario, 
like in \rf{dyn},
has been undertaken in the gravity sector \cite{b21}.
Finally we note some recent theoretical work 
has identified general properties of backgrounds 
in effective field theory \cite{kl21}, 
and new types of tests are possible that search for non-Riemann geometry \cite{Kostelecky:2021tdf}.

Of the two approaches identified above, 
the latter, 
equation \rf{lag2},
is appropriate for short-range gravity tests.
Such tests involve weak 
gravitational fields 
in the Earth laboratory setting,
thus the typical size of components of $h_\mn$ are much less than unity, 
in cartesian coordinates.
Furthermore, 
to keep a reasonable scope we will
truncate the series \rf{sqk} to mass dimensions $4$, $5$, and $6$.

Any study of actions with higher than second order derivatives is subject to well-known results, 
such as Ostragradsky instabilities \cite{Woodard:2015zca}.
In the present paper, 
while the test framework \rf{lag2} is viewed perturbatively, 
with the higher derivative terms 
as small corrections \cite{km09}, 
our discussion of solutions beyond leading order in coefficients will overlap with features in higher derivative models.
Some features are discussed in our results in Section \ref{isotropic Newtonian limit, exact solutions} and Section \ref{anisotropic exact solutions}.

\subsection{Prior short-range gravity results}
\label{prior short-range gravity results}

In references \cite{bkx15} and \cite{km17},
Lorentz-symmetry breaking solutions
for short-range gravity tests were found using an approximation of first order 
in the coefficients.
We summarize these results briefly here for comparison.
Assuming a static matter source and using the framework of \rf{lag2}, 
one solves the field equations perturbatively assuming
any modifications to the field equations from symmetry-breaking terms are small \cite{bk06, bkx15, km17}.
The leading order modified Newtonian potential from a point mass $m$ at the origin can be written in terms
of Newton spherical coefficients $k_{jm}^{N(d)lab}$ as a series
\beq
U = \fr {G_N m}{r} 
+ \sum_{djm} \fr {G_N m}{r^{d-3}} 
\, Y_{jm} (\th, \ph) k_{jm}^{N(d)lab},
\label{potpert}
\eeq
where the angular dependence $\th, \ph$ in the spherical harmomics $Y_{jm} (\th, \ph) $ pertains to the vector from the origin to the field point 
$\vec r = r (\sin \th \cos \ph, \sin \th \sin \ph, \cos \th )$ and $r=|\vec r|$.
The spherical coefficients
$k_{jm}^{N(d)lab}$ are related to the coefficients in Eq.\ \rf{sqk} as linear combinations, 
but the expressions are lengthy
and omitted here, 
and relations between the dimension label $(d)$ and the allowed values of $j$ can be found in \cite{km17}.
The superscript ``lab" means that the coefficients are written in the laboratory coordinate system.
Typically, 
the lab frame coefficients are re-expressed in terms of the Sun-centered Celestial Equatorial Frame coefficients using an observer Lorentz transformation, 
revealing harmonic time dependence
\cite{kl99,km02,kv15}.

The result in equation \rf{potpert} has already been used for analysis in experiments \cite{Long:2014swa,Shao:2015gua,Shao:2016jzh,Shao:2018lsx}.
In fact, 
new experiments can be designed to maximize the type of anisotropic signal in \rf{potpert} \cite{Shao:2016jzh,Chen:2017bru,Bobowski:2022qqa}.
Recent result place limits on $14$ $k_{jm}^{N(6)}$ coefficients
and $22$ $k_{jm}^{N(8)}$ coefficients at the $10^{-9} \, m^2$
and $10^{-12} \, m^4$ levels, 
respectively.
However, 
the leading order approximation used for \rf{potpert} makes searches in some short-range tests challenging, 
as some tests are designed 
to probe very small length scales 
at the cost of sensitivity to the 
Newtonian force from the test masses \cite{Decca:2005qz}.
Such tests often lie outside the range of applicability of the result \rf{potpert}, 
which assumes the extra correction term to the Newtonian potential is smaller than 
the first term.

One other observation is that, 
with the exception of mass dimension $4$ coefficients, 
no rotational scalar coefficients, 
or isotropic coefficients show up in the result \rf{potpert}.
In fact, 
it has been shown that one combination of isotropic coefficients does show up in the perturbative analysis, 
but only as contact term that vanishes outside of the matter distribution \cite{bkx15}.
As we show below, 
a non-perturbative treatment reveals 
in more detail the role played by these coefficients.

\section{Isotropic coefficients, Newtonian limit, nonperturbative}
\label{isotropic Newtonian limit, exact solutions}

\subsection{Special case model}
\label{Special case model}

We begin with a special case to illustrate the features of the solutions studied in this work.
One particular model that contains the interesting features of exact short-range solutions is the following Lagrange density:
\beq
\cL = \frac {1}{2\ka} \sqrt{-g} \left( R + k_{\al\be} R^{\al\be} R \right), 
\label{action1}
\eeq
which is a special case of \rf{lag1}.
The second term is the non-standard one
with the coefficients for Lorentz violation denoted $k_{\al\be}$.
These $10$ quantities have units of length squared or inverse mass squared in natural units.

The action in \rf{action1}, 
yields the field equations in appendix \rf{k2case}, 
upon variation with respect to the full metric $g_\mn$.
In the linearized gravity limit, 
and assuming the coefficients $k_\ab$
have vanishing partials $\prt_\al k_{\be\ga}=0$, 
the field equations \rf{k2case} become,
\bea
(G_L)^\mn &=& -\frac 12 \et^\mn k_\ab \prt^\al \prt^\be R_L
-\et^\mn k_\ab \prt^\ga \prt_\ga  (R_L)^\ab 
+\frac 12 k^{\nu \al} \prt_\al \prt^\mu R_L
\nonumber\\
&&
+\frac 12 k^{\mu \al} \prt_\al \prt^\nu R_L
+ k_\ab \prt^\mu \prt^\nu (R_L)^\ab
-\frac 12 k^\mn \prt^\al \prt_\al R_L + \ka T^\mn,
\label{k2linear}
\eea
where $T^\mn$ is the matter stress-energy tensor.
Note that in the linearized gravity case, 
indices are raised and lowered with
$\et_\mn$,
the linearized Ricci tensor is
$(R_L)_\mn = (1/2)(\prt_\mu \prt^\al h_{\al\nu}+\prt_\nu \prt^\al h_{\al\mu} -\prt_\al \prt^\al h_\mn-\prt_\mu \prt_\nu h^\al_{\pt{\al}\al})$,
$R_L= \prt^\al \prt^\be h_{\al\be} - \prt_\al \prt^\al h^\be_{\pt{\be}\be}$, 
and $(G_L)^\mn  = (R_L)^\mn - (1/2) \et^\mn R_L$.
The task is next to obtain a space and time component decomposition of these field equations \rf{k2linear}.

If we further restrict attention 
to the static limit and only isotropic coefficients $k_{00}$ and $k_{jj}$, 
in a special coordinate system,
we obtain the following coupled equations
for the metric components $h_{00}$ and $h_{jj}$ (in harmonic gauge):
\bea
\nabla^2 ( h_{00} + h_{jj} )
-3 (\koo- \frac {1}{9} \kjj ) \nabla^4 h_{00} 
+ (\koo - k_{ll} ) \nabla^4 h_{jj} &=& 
-32 \pi G_N \rh,
\nonumber\\
 \nabla^2 (3 h_{00} + h_{jj} ) 
+ 4 (\koo - \frac{1}{3} k_{ll} ) \nabla^4 h_{00} 
+ \frac {8}{3} k_{ll} \nabla^4 h_{jj}
&=& 0.
\label{coupled}
\eea
Note that $\kjj - \koo= k_\mn \et^\mn$ is a Lorentz invariant scalar
combination. 
We have assumed a static pressure-less matter distribution so that only 
$T^{00}=\rh$ is nonzero in $T^\mn$.
We also find in this limit that the equation for $h_{0i}$ is simply Laplace's equation:
\beq
\nabla^2 h_{0i} = 0.
\label{h0i}
\eeq

For the remaining components of $h_{ij}$ it is advantageous to express the solution in terms of a traceless piece.
By this we mean that if the equations for $h_{ij}$
are denoted ${\cal E}_{ij}=0$,  
the relevant projection is ${\cal E}_{ij}- (1/3) \de_{ij} {\cal E}_{kk}$.
This yields
\beq
\nabla^2 ( h_{ij}-\frac{1}{3} \de_{ij} h_{kk} ) 
- 2 (\koo - \frac{1}{3} \kjj ) \cD_{ij} \nabla^2 h_{00}
- \frac{4}{3} \kjj \cD_{ij} \nabla^2 h_{kk}=0,
\label{offdiag}
\eeq
where $\cD_{ij}=\prt_i \prt_j - \frac{1}{3} \de_{ij} \nabla^2$ is a traceless operator.  
Evidently, 
if one can solve independently for $h_{00}$ and $h_{ll}$, 
then equation \rf{offdiag} can be viewed as an inhomogeneous
equation for the traceless piece of $h_{ij}$ with source terms 
involving projections of $h_{00}$ and $h_{ll}$.


Our main focus is to solve the equations \rf{coupled} for $h_{00}$ and $h_{jj}$, 
since $h_{00}$ is the metric component directly 
related to the Newtonian potential $U_N$ via $h_{00}=2 U_N$.
The solution can be found using standard methods of solving PDEs.
We first discuss the construction of 
a Green function solution where 
we assume a point source $4\pi G_N \rh = \de^{(3)}(\vec r - \vec r^\prime )$.
The point source solutions for $h_{00}$ and $h_{jj}$ are denoted $\cG_1$ and $\cG_2$.

Given the form of the solution to the 
equation with $\nabla^2$ and $\nabla^4$ in appendix \rf{nl1}, 
we propose the ansatz that the general solutions will take the form of the following functions of $R=|\vec r- \vec r^\prime |$:
\bea
    \cG_1 = {1 \over R } \left( A_1 e^{-q_{1} R}+A_{2} e^{-q_{2} R}+A_{3}\right),
    \nonumber\\
    \cG_2 = {1 \over R } \left( B_ 1 e^{-q_{1} R}+B_{2} e^{-q_{2} R}+B_{3}\right),
    \label{newpot}
\eea
Here the $A_n$'s and $B_n$'s are constants to be solved for as well as the $q_1$ and $q_2$.  
In constructing this solution we are assuming the boundary conditions such that the metric components go to zero far from the source, 
and we neglect any homogeneous solutions
to \rf{coupled}.
Insertion of \rf{newpot} into the point source version of \rf{coupled}, 
followed by using the properties of functions of $R=|\vec r- \vec r^\prime |$, 
allows one to solve for the $8$ parameters 
$A_1$, $A_2$, $A_3$, $B_1$, $B_2$, $B_3$, $q_1$, and $q_2$
from $8$ resulting algebraic equations.

First, 
we find that for nontrivial solutions, 
both $q_1^2$ and $q_2^2$ must satisfy the 
quartic equation:
\beq
1+ (\koo - \frac {5}{3} \kjj ) q^2 + ( \koo + \frac {1}{3} \kjj )^2 q^4 =0.
\label{quartic}
\eeq
The solutions to \rf{quartic} can be obtained from the quadratic result,
\beq
q^2 = u \pm v,
\label{Qsoln1}
\eeq
where $u$ and $v$ are given by:
\bea
u &=& { - (\koo - \frac 53 \kjj) \over 2 (\koo + \frac 13 \kjj )^2 },
\nonumber\\
v &=& { \sqrt{(\koo - \frac 53 \kjj )^2 - 4 (\koo + \frac 13 \kjj )^2} \over 
2 (\koo + \frac 13 \kjj )^2 }
\label{uv}
\eea
The four possible roots of the equation \rf{quartic} can be obtained generally 
by taking the complex square roots of \rf{Qsoln1}. 
The position of $z=q^2$ in the complex plane depends on the values of the coefficients
$\koo$ and $\kjj$.
Note that $u$ is real and $v$ can be real or complex.
The values of the coefficients determine the properties of the $4$ possible roots $\{ q=z^{1/2}= \pm (u \pm v)^{1/2} \}$.
If $q$ is entirely real and positive, 
then the solutions in \rf{newpot} will exhibit exponential damping in $R$ 
or short-range Yukawa-like behavior.
The case where $q$ is negative and real will result in runaway exponential increase and is not physically viable.
When $q$ has an imaginary piece or is entirely imaginary, 
the solution will have oscillations in $R$.

In what follows we assume the 
condition $q_1^2 \neq q_2^2$.  
This condition ensures that the coefficients $\koo$ and $\kjj$ are treated {\it a priori} independent.
This condition implies that in \rf{Qsoln1}, 
$q_1^2$ takes one sign in the $\pm$, 
and $q_2^2$ takes the other sign.
For this case we obtain the solutions for the Green function $\cG_1$ as follows.
\bea
\cG_1 &=& {1 \over 2\pi R}  
- {1 \over 4 \pi } \left( 1 + { \koo + \frac {11}{3} \kjj \over 
\sqrt{(\koo - \frac 53 \kjj )^2 - 4 (\koo + \frac 13 \kjj )^2} } \right)
{ e^{-R/\la_+} \over R} 
\nonumber\\
&&
- {1 \over 4 \pi } \left( 1 - { \koo + \frac {11}{3} \kjj \over 
\sqrt{(\koo - \frac 53 \kjj )^2 - 4 (\koo + \frac 13 \kjj )^2} } \right)
{ e^{-R/\la_-} \over R},
\label{Greensoln1}
\eea
where the $\la_\pm$ constants are defined by
\beq
{ 1 \over (\la_{\pm})^2 } = u \pm v,
\label{lambdas}
\eeq
and they act like two distinct length scales.

We note the contrast of this result with previous results.
First, unlike the Yukawa potential, 
\beq
U_{Y} = \fr {Gm}{r} \left( 1 + \al e^{-r/\la} \right),
\label{yukawa}
\eeq
we have $2$ length scales in \rf{lambdas}.
Second, 
the amplitudes of the two terms vary depending on the values of the coefficients.
In particular, 
we find that these amplitudes could take on large values for a narrow range of coefficient ratios $\kjj / \koo$, even
if the coefficients themselves are small compared to the length scales probed.
This is in contrast to standard assumptions of the smallness of Lorentz-violating effects.  
Note that the length scales would also be small, 
so such large Lorentz-breaking forces could
escape detection in long-range tests, 
and this philosophy is along the lines of proposals for new short-range forces more generally.

\begin{figure}[h!]
\begin{center}
\includegraphics[width=4in]{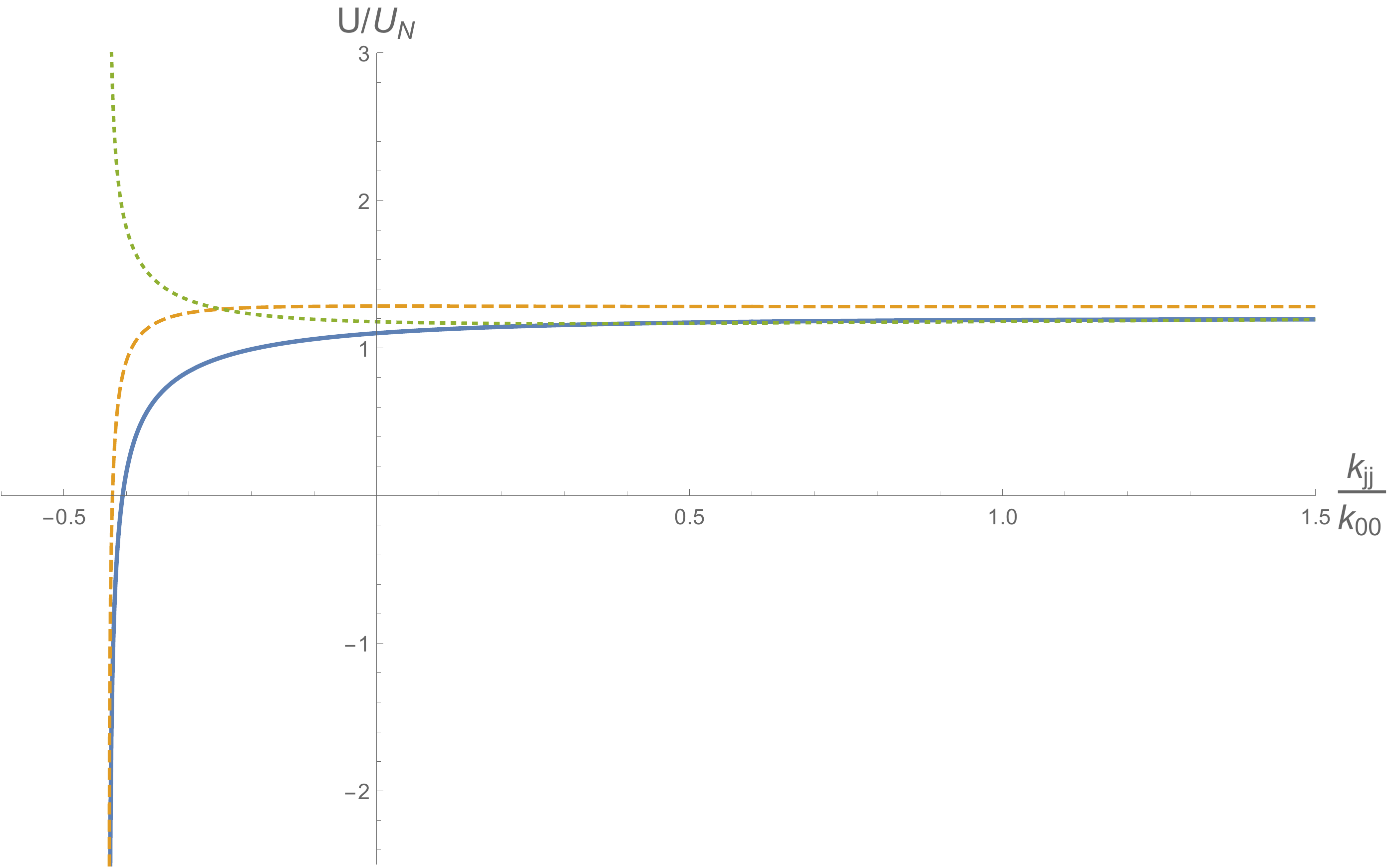}
\caption{\label{samplecasepotplot} A plot of the ratio of the modified potential of equation \rf{Greensoln1}
divided by the Newtonian potential
for a unit mass at the origin.
The horizontal axis is the coefficient 
ratio $k_{jj}/\koo$ while the solid curve, 
dashed curve, 
and dotted curves are
the cases of different positions away from the point mass, 
$R=\sqrt{k_{00}}$,
$R=3\sqrt{k_{00}}$,
and $R=5\sqrt{k_{00}}$, 
respectively.
}
\end{center}
\end{figure}

To get an idea of the behavior of these solutions as the values of the coefficients change, 
in figure \ref{samplecasepotplot},
we plot the potential $U=2\cG_1$ 
for a point mass of unit strength
as a function of $\kjj / \koo$ for several values of the distance $R$
(more specifically, the ratio
of the distance $R$ to $\sqrt{k_{00}}$).
This can be compared to the standard Newtonian potential which would be a horizontal line in the same graph.
We clearly see a singular point in the $\kjj$, $\koo$ parameter space as the $\kjj/\koo$ approaches $-3/7$.

The solution obtained in \rf{Greensoln1} above agrees precisely with an alternative method, 
where one uses Fourier decomposition in momentum space, 
followed by contour integration.
For practical evaluation over distributions of matter, 
such as those used in experiment, 
one would take the integral of the Green functions
over the smooth matter distributions 
$\rh(\vec r^\prime)$ as usual.
Thus, 
since $h_{00}=2U$,
the Newtonian potential is
\beq
U= 2 \pi G_N \int d^3r^\prime 
\rh (\vec r^\prime) 
\cG_1 (\vec r, \vec r^\prime ).
\label{Newt}
\eeq

\subsection{General effective field theory case}
\label{general EFT case}

Here we consider generalizing the solution of section \ref{Special case model}.
A more general treatment includes 
the quadratic Lagrange density of \rf{lag2}.
First we examine the field equations for this approach, 
which are obtained from \rf{lag2} by variation with respect to $h_\mn$:
\beq
     (G_L)^{\mu\nu} 
     + \de M^{\mu\nu\rh\si} h_{\rh\si}
     =8 \pi G_N T^\mn,
\label{eom1}
\eeq
where we have adopted the notation of 
Ref.\ \cite{km17} with
\bea
\de M^{\mu\nu\rh\si} h_{\rh\si}&=&
-\Big[ \frac{1}{4}(\hat{s}^{\mu\rh\nu\si}
     +\hat{s}^{\mu\si\nu\rh})
     +\frac{1}{2}\hat{k}^{\mu\nu\rh\si} 
     \nonumber\\
&&  \pt{+} +\frac{1}{8}(\hat{q}^{\mu\rh\nu\si}
     +\hat{q}^{\nu\rh\mu\si}
     +\hat{q}^{\mu\si\nu\rh}
     +\hat{q}^{\nu\si\mu\rh}) \Big] h_{\rh\si}.
\label{deM}
\eea

\subsubsection{Determining the field equations}
\label{determining the field equations}

Next we will focus attention on mass dimension $d=6$ or less to keep the scope reasonable.
Furthermore, 
as we are taking the static limit, 
as the prior section, 
only spatial derivatives appear.
We will again choose to look at only isotropic coefficients, 
as these we expect to result 
in field equations
we can solve exactly 
in analytic form, 
and to reveal the role of these 
coefficients in short-range gravity tests.

It is not exactly trivial to extract the isotropic
coefficients in the expansions \rf{sqk}
but we outline the process here 
and leave most of the details for the appendix.
Consider the $\mu=0,\nu=0$ component of \rf{deM}, 
including terms up to mass dimension $6$.
Tensor symmetry properties of the coefficients in
\rf{deM} can be used to eliminate some contributions
outright (see the Young Tableau in Table 1 of Ref.\ \cite{km16} and the  appendix of this paper \rf{spacetimeyoung} ).
For instance, 
antisymmetry of the indices yields ${\hat s}^{0000}=0={\hat s}^{000i}$ and ${\hat q}^{000i}=0$.
The surviving contributions to the $00$ component
of \rf{deM} are initially collected as
\bea
\de M^{00\rh\si}h_{\rh\si} &=& 
-\frac 12 k^{(6)0i0j0k0l} \prt_{ijkl} h_{00}
\nonumber\\
&&
\hspace{-10pt}
-\{
\frac 12 k^{(6)0j0k0lim} \prt_{jklm} 
+ \frac 14 q^{(5)0ij0k0l} \prt_{jkl} 
\} 
h_{0i}
\nonumber\\
&&
\hspace{-10pt}
-\{ 
\frac 12 [
s^{(4)0ik0jl} \prt_{kl} + s^{(6)0ik0jlmn} \prt_{klmn} +k^{(6)0k0limjn} \prt_{klmn} ]
\nonumber\\
&&
\pt{-- }+ \frac 14 [q^{(5)0ik0ljm} + q^{(5)0jk0lim}] \prt_{klm} 
\}
h_{ij}
\label{deM00_1}
\eea
where we make use of a short-hand ($\prt_{ijk...} = \prt_i \prt_j \prt_k ...$) 
for multiple partials.
Among the coefficients occurring in \rf{deM00_1}, 
those that are isotropic will be invariant under
observer rotations $SO(3)$, 
and thus expressible in terms of rotational scalar contractions, the kronecker delta $\de^{ij}$ and
the levi-civita $\ep^{ijk}$. 

As an example of how to decompose the terms in \rf{deM00_1}, 
consider the first term on the first line with the coefficients $k^{(6)0i0j0k0l}$, 
which has $ij$ and $kl$ index symmetry.
This would lead us to the only available scalar contractions being $k^{(6)0i0i0j0j}$ and $k^{(6)0i0j0i0j}$.
However, 
because the underlying tensor satisfies
the cyclic identity
$k^{(6)\mu\ep_1\nu\ep_2\rh\ep_3\si\ep_4}
+k^{(6)\mu\ep_1\nu\ep_3\ep_2\rh\si\ep_4}
+k^{(6)\mu\ep_1\nu\rh\ep_2\ep_3\si\ep_4}=0$, 
one can show that 
$k^{(6)0i0j0i0j} = k^{(6)0i0i0j0j}$.
Therefore the $k^{(6)0i0j0k0l}$
coefficients, 
in the isotropic limit, 
must be proportional to combinations
of kronecker deltas $\de^{ij} \de^{kl}+...$
and the one scalar $k^{(6)0i0i0j0j}$.
Symmetry considerations lead us to
\beq
k^{(6)0i0j0k0l} = \frac {1}{15} (\de^{ij} \de^{kl}
+\de^{ik} \de^{jl}
+\de^{il} \de^{jk})  k^{(6)0m0m0n0n},
\label{k60i0j0k0l-1}
\eeq
and thus
\beq
-\frac 12 k^{(6)0i0j0k0l} \prt_{ijkl} h_{00} = - \frac {1}{10} k^{(6)0m0m0n0n} \nabla^4 h_{00}.
\label{k60i0j0k0l-2}
\eeq
which simplifies the first term in \rf{deM00_1}
to the desired form.

For the second line in \rf{deM00_1}, 
the coefficients $k^{(6)0j0k0lim}$ and 
$q^{(5)0ij0k0l} $ do not appear to have 
any scalar contractions due to the number of indices, 
or the symmetry properties.
Nor can they be written in terms of purely $\de_{ij}$
and $\ep_{ijk}$.
We conclude their isotropic limit contribution vanishes:
\beq
-\{ \frac 12 k^{(6)0j0k0lim} \prt_{jklm} 
+ \frac 14 q^{(5)0ij0k0l} \prt_{jkl} 
\} 
h_{0i} \stackrel{\mathrm{iso}}{=} 0. 
\label{k6jklim}
\eeq
One proceeds along similar lines for the remaining terms in 
\rf{deM00_1}.
The details are relegated to the appendix.

The final simplification to the isotropic coefficient case for \rf{deM00_1} results in
\bea
\de M^{00\rh\si}h_{\rh\si} &=& 
- \frac {1}{10} k^{(6)0m0m0n0n} \nabla^4 h_{00}
\nonumber\\
&&
+\{ 
\frac {1}{12} s^{(4)0kl0kl} 
+\frac {1}{30} s^{(6)0kl0klmm} \nabla^2  
+\frac {1}{30} k^{(6)0k0klmlm} \nabla^2 
\} 
\nonumber\\
&&
\times
(\prt_i \prt_j - \de^{ij} \nabla^2 ) h_{ij}.
\label{deM00_2}
\eea
Note the absence of the $h_{0i}$ components
in this case.
The remaining components of \rf{deM} $\mu=0,\nu=i$
and $\mu=i,\nu=j$
are worked out in the appendix.
The equation for $h_{0i}$ decouples from the remaining components of $h_\mn$ and we display below the coupled equations for the components
$h_{00}$ and $h_{jj}$.
As in the special case of the previous section, 
the off-diagonal components of $h_{ij}$ can be obtained from a traceless version of appendix \rf{deMs_2}, 
of secondary interest in this work.

To obtain the relevant differential equations we make 
the partial choice of gauge:
$\prt_j h_{ij} = \prt_i (h_{jj} - h_{00})/2$.
Furthermore, 
it will be convenient for solving
the differential equations to work with 
the trace-reversed components
$\hb_{00}=(1/2)( h_{00}+ h_{jj} )$ 
and $\hb_{jj} = (1/2) (3 h_{00} - h_{jj} )$.
Also, 
since they can be probed with other tests \cite{Hees:2016lyw}, 
we disregard the mass dimension $4$ isotropic coefficients.
With these choices and the results of the appendix, 
the two coupled equations are given by,
\bea
-\frac 12 [ \nabla^2 + (k_1+k_2) \nabla^4 ] \hb_{00}
-\frac 12 k_1 \nabla^4 \hb_{jj} &=& 8\pi G_N \rh,
\\
-\frac 12 [ \nabla^2 + k_2 \nabla^4 ] \hb_{jj}
-\frac 12 (k_2+ k_3) \nabla^4 \hb_{00} &=& 0,
\label{SMEcoupled1}
\eea
where the $k_1$, $k_2$, and $k_3$ are the combinations
\bea
k_1 &=& \frac {1}{10} k^{(6)0i0i0j0j},
\nonumber\\
k_2 &=& \frac {1}{15} [ s^{(6)0kl0klmm} + k^{(6)0k0klmlm} ],\nonumber\\
k_3 &=& \frac {1}{18} s^{(6)klmklmnn}
+ \frac {1}{15} k^{(6)klklmnmn}.
\label{constants}
\eea
These equations are very similar to 
those in \rf{coupled}, 
except that now we have $3$ {\it a priori} independent combinations of coefficients, 
instead of $2$.
The combinations appearing in \rf{constants}
overlap with the isotropic coefficient combination appearing in GW tests, 
which is in appendix \rf{k00}.

It is important to emphasize 
that the assumption of isotropy in a special coordinate system is a special case of the general coefficients
in the EFT framework.
The focus here is on these particular coefficients, 
effectively setting the others to zero.
However, 
in principle one can use the coordinate covariance of the EFT to transform the coefficients from one frame
to another.
Isotropic coefficients are rotational scalars.
Under $SO(3)$ rotations of the spatial 
coordinates $(x^\prime)^j=R^j_{\pt{j}k} x^k$
they do not change.
Under observer boosts, 
however, 
the components would mix with others.
Once one introduces a boost velocity $\be$,
this is typically of order $10^{-4}$, 
and to be consistent one needs the full post-Newtonian metric with includes
the velocity of matter $v^j$ included.
We do not consider this here but it has been done elsewhere for coefficients in the gravity sector \cite{bk06,kt11,bh17}.

\subsubsection{Solving the coupled equations}
\label{solving the coupled equations}

With a similar approach to section \ref{Special case model}, 
we seek Green function solutions
for a unit point source $ 
4\pi G_N \rh = \de^{(3)}(\vec R)$, 
and choose boundary conditions so that
the fields vanish at spatial infinity.
Denoting the Green functions for $\hb_{00}$
and $\hb_{jj}$ as $G_1$ and $G_2$, 
respectively, 
we obtain the Green function matrix equation,
\bea
\left( 
\begin{array}{cc} 
\nabla^2 + (k_1 + k_2) \nabla^4 & 
k_1 \nabla^4  \\
(k_2 + k_3 ) \nabla^4  & 
\nabla^2 + k_2 \nabla^4   \\
\end{array}
\right) 
\left( 
\begin{array}{c} 
G_1 \\
G_2 \\
\end{array}
\right)= 
\left( 
\begin{array}{c} 
-4 \de^{(3)} (\vec R) \\
0 \\
\end{array}
\right).
\label{SMEGreenEqn1}
\eea
Next we use Fourier transforms of the Green functions via
\beq
G_n = \frac {1}{(2\pi)^3} 
\int d^3p e^{i\vec p \cdot \vec R} \tG_n,
\label{Fourier}
\eeq
where $n=\{ 1,2 \}$.
With this, 
the matrix equation becomes algebraic in momentum space as
\bea
\left( 
\begin{array}{cc} 
-1 + (k_1 + k_2) p^2 & 
k_1 p^2  \\
(k_2 + k_3 ) p^2  & 
-1 + k_2 p^2  \\
\end{array}
\right) 
\left( 
\begin{array}{c} 
p^2 \tG_1 \\
p^2 \tG_2 \\
\end{array}
\right)= 
\left( 
\begin{array}{c} 
-4 \\
0 \\
\end{array}
\right),
\label{SMEGreenEqn2}
\eea
where $p=|\vec p|$ and $p^2$ is factored from the matrix to make it unitless.
Since it is of crucial importance for the pole structure and 
the solutions, 
we record here the determinant of the matrix $M$ in \rf{SMEGreenEqn2}:
\beq
\det M = 1 - (k_1 + 2 k_2) p^2 + (k_2^2 - k_1 k_3 )p^4.
\label{detM}
\eeq

Inverting the matrix in \rf{SMEGreenEqn2}, 
we obtain the momentum space solutions:
\bea
\tG_1 &=& \fr {4 (1-k_2 p^2)}{p^2 \det M},
\nonumber\\
\tG_2 &=& \fr {4 (k_2 + k_3)}{\det M}.
\label{GreensolnMom}
\eea
Inserting the results into \rf{Fourier}, 
and taking advantage of the spherically symmetric nature of the solutions in \rf{GreensolnMom}, 
we can directly integrate the angular part via $d^3p = p^2 dp d \Om_p$, 
What remains is a one-dimensional Fourier
transform integral over the magnitude of the momentum $p$.
For instance, 
for $G_1$ we obtain
\beq
G_1 = -\fr {i}{\pi^2 R} \int_{-\infty}^{\infty} e^{i p R} \fr {1-p^2 k_2}
{p[1 - (k_1 + 2 k_2) p^2 + (k_2^2 - k_1 k_3 )p^4]} dp,
\label{G1}
\eeq
with a similar integral for $G_2$.
This integral may be evaluated using 
contour integration in complex $p$ space.
Clearly the poles of \rf{detM} play a strong role.  

The result of the complex integration 
calculation gives the Green functions $G_1$ and $G_2$ in position space.
We find,
\bea
G_1 &=& \fr {1}{\pi R} \left[ 
1 + \frac 12 \left( \fr { \ze_1 \ze_\al }{\sqrt{1+4\ch}}
-1 \right) e^{\pm i w_1 R}
- \frac 12 \left( \fr { \ze_1 \ze_\al }{\sqrt{1+4\ch}}
+1 \right) e^{\pm i w_2 R} 
\right],
\nonumber\\
G_2 &=& \fr { \ch \ze_1 \ze_\al }{ \sqrt{1+4\ch} \pi R} (e^{\pm i w_1 R} - e^{\pm i w_2 R} ),
\label{GsFin}
\eea
where we define $\ch$, the poles
$w_1$ and $w_2$, and the ``zetas" as
\bea
\ch &=& \fr {k_2+k_3}{k_1},
\\
w_1 &=& \fr {1}{\sqrt{2 | k_2^2-k_1 k_3|} } 
\left( \ze_\al (k_1+2k_2) + |k_1| \sqrt{1+4\ch} \right)^{1/2},
\\
w_2 &=&\fr {1}{\sqrt{2 | k_2^2-k_1 k_3|} } 
\left( \ze_\al (k_1+2k_2) - |k_1| \sqrt{1+4\ch} \right)^{1/2},\\
\ze_1 &=& {\rm sign} (k_1), \\
\ze_\al &=& {\rm sign} (k_2^2 - k_1 k_3).
\label{quants}
\eea
The $\pm$ signs in the exponential functions 
are to be chosen to ensure an exponential decay rather than growth, 
and the choice depends on the sign 
of the complex part of $w_1$ and $w_2$.

Examination of the solutions \rf{GsFin} reveals that the amplitudes of the exponential terms appear to become arbitrarily large as $\ch \rightarrow -1/4$
from above or below.
However, 
in the same limit we have $w_1$ appearing to coincide with $w_2$, 
and so the two terms in \rf{GsFin} appear as though
they might cancel.
So it is not immediately clear 
the behavior of the solution in this limit.
To understand the general solution better, 
we explore some limiting cases.

\subsubsection{Exploration of solutions}
\label{exploration of solutions}

First, 
we will focus the attention 
on the combination of the Green functions related to the Newtonian potential, $\cG_1 = (1/2)( G_1 +G_2 )$.
This simplifies to
\beq
\hspace{-30pt}
\cG_1 = \fr {1}{2 \pi R} \Bigg[ 
1 + \fr 12 \left( \fr { \ze_1 \ze_\al (1+2\ch) }{\sqrt{1+4\ch}}
-1 \right) e^{\pm i w_1 R}
- \fr 12 \left( \fr { \ze_1 \ze_\al (1+2 \ch )}{\sqrt{1+4\ch}}
+1 \right) e^{\pm i w_2 R} 
\Bigg].
\label{cG1}
\eeq
Note that this solution reduces to the one appearing in the prior section \ref{Special case model} with the
substitutions $k_1=\kjj/3 - \koo$,
$k_2=\koo-\kjj$, and $k_3=8\kjj/3$.
The Newtonian potential for a realistic source is obtained from the matter distribution integral \rf{Newt}.

Consider a sample case of the $k_1$, $k_2$, 
and $k_3$ parameter space.
Let $\ch=-6/25$ so that $\sqrt{1+4\ch} = 1/5$.
Then we further specialize to the case $k_3=0$.
Inserting these assumptions into \rf{cG1} leaves a solution valid for a one parameter subset (chosen as $k_2$).
Specifically we find
\bea
\cG_1 = 
\left\{ 
\begin{array}{cc}
\fr {1}{2 \pi R} \left( 
1 - \fr {9}{5}  e^{-R/\la_1} + \fr {4}{5} e^{-R/\la_2}
\right), & k_2 >0 \\
\fr {1}{2 \pi R} \left( 
1 - \fr {9}{5}  e^{\mp iR/\la_1} + \fr {4}{5} e^{\mp i R/\la_2}
\right), & k_2 <0 \\
\end{array}
\right.
\label{sample1}
\eea
where the length scales are 
$\la_1=\sqrt{3 |k_2|/2}$
and $\la_2=\sqrt{2 |k_2|/3}$.
Note that in this case, 
with a negative sign for $k_2$, 
one obtains purely oscillatory corrections 
with no exponential damping.
For this latter case, 
if desired one can obtain a real solution by superposition of the two signs. 

If the length scale of the coefficients, $\la \sim \sqrt{|k_2}|$, 
are expected to be small compared to 
accessible laboratory length scales than
the solution for $k_2>0$ is consistent
with a new force that arises only on short scales.
This situation is consistent with the 
spacetime symmetry breaking being small, 
and the terms added to the action being small 
corrections to known physics.
On the other hand, 
if $k_2<0$,
with small length scales $\la$, 
one finds a rapidly spatially varying 
Newtonian potential with a substantial
(of order unity or higher) amplitude.
The lack of such observed long range forces
could be used to theoretically reject
this region of the coefficient space 
of solutions as unphysical. 
Note that this latter case bears 
similarity to considerations of higher derivative models where, 
in some cases,
one does not find a smooth limit to a perturbative approach \cite{Simon:1990ic,Eliezer:1989cr}.
Similarly here, 
trying extrapolate
$k_2 \rightarrow 0$ when $k_2<0$ 
simply results in rapidly varying (unobserved) forces.
In contrast, 
again, 
the former solution $k_2>0$ 
with the decaying exponential reduces to
a delta function at the origin when $k_2 \rightarrow 0$, 
like the contact term found by a perturbative approach in Ref.\ \cite{bkx15}.
Such terms also arise in other models \cite{Kostelecky:2021xhb}.

\begin{figure}[h!]
\begin{center}
\includegraphics[width=4in]{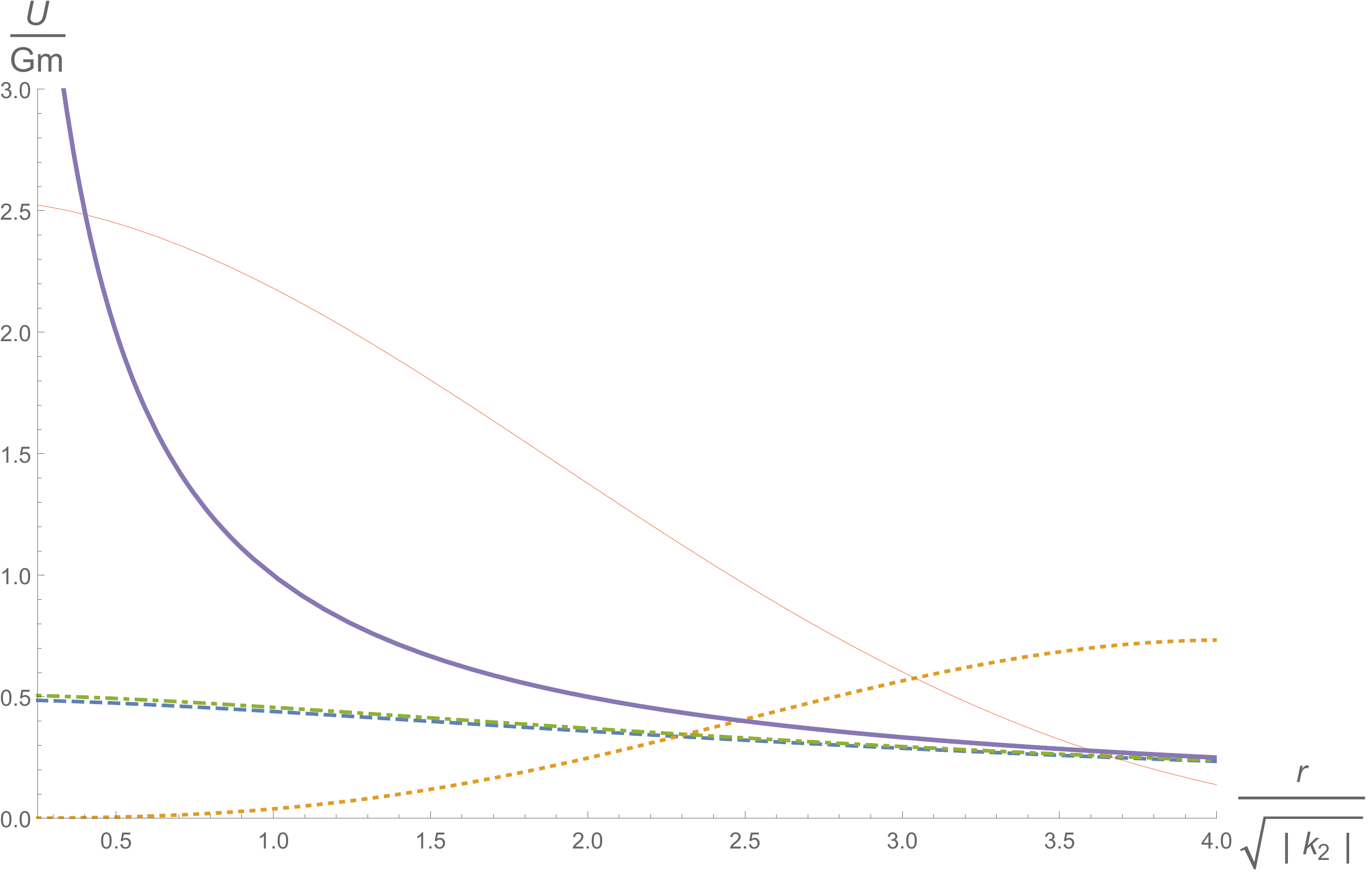}
\caption{\label{SMEpotplot} A plot of the modified potential of equations \rf{sample1} and \rf{sample2}
for a unit mass at the origin ($r=0$).
The vertical axis is the scaled potential and
the horizontal axis is the distance from the source in terms of the length unit $\sqrt{|k_2|}$.
The solid curve is the Newtonian potential $1/r$ falloff.
The dashed and dotted curves are
the solutions in equation \rf{sample1},
with $\ch=-6/25$ and
$k_2>0$ and $k_2<0$, 
respectively.
The dot-dashed and thin curves
are the solutions from equation 
\rf{sample2} for $\ch=-13/50$, 
again with $k_2>0$ and $k_2<0$, 
respectively.}
\end{center}
\end{figure}

Next we look at what happens 
when we approach $\ch=-1/4$ from ``below".
Consider $\ch=-13/50$ so that $\sqrt{1+4\ch} = i/5$.
Again we assume $k_3=0$ and we obtain in this case,
\bea
\cG_1 = 
\left\{ 
\begin{array}{cc}
\fr {1}{2 \pi R} \left( 
1 - e^{-R/\la_1} [\cos (R/\la_2)+\frac{12}{5} \sin (R/\la_2)] 
\right), & k_2 >0 \\
\fr {1}{2 \pi R} \left( 
1 - e^{-R/\la_2} [\cos (R/\la_1)-\frac{12}{5} \sin (R/\la_1)] 
\right), & k_2 <0 \\
\end{array}
\right.
\label{sample2}
\eea
where now $\la_1 = \sqrt{26|k_2|/25}$
and $\la_2 = \sqrt{26|k_2|}$.
We now have a damped exponential behavior 
accompanied with oscillatory behavior in $R$.
Changing the sign of $k_2$ merely swaps 
the length scales involved in damping 
versus oscillations.
We plot the cases described above in \rf{sample1} and \rf{sample2}
in Figure \ref{SMEpotplot}.
All of the examples exhibit behavior strikingly different from the Newtonian case.
Note in particular, 
a resemblance of the modified Newtonian potential 
solutions in \rf{sample2} to a Dilaton-gravity coupling proposed long ago \cite{Fujii:1971vv}.

Returning to the general case of \rf{cG1}, 
we enumerate the different functional 
forms of the solution for different regions of
the space spanned by the coefficients $k_1$, 
$k_2$, 
$k_3$
in Table \ref{solntable1}. 
We assume that we do not make contact with 
the singular point in $w$ \rf{quants}, 
$k_2^2=k_1 k_3$, 
as this point would correspond to the dissappearance
of the $p^4$ term in \rf{detM}, 
and would impose a condition on the 
{\it a priori} independent coefficients.
Nonetheless we include this case in the appendix \ref{Special cases} since it may be of interest.
Also, 
as in section \ref{Special case model}, 
we can explore what happens near the apparent
singularity in the solution \rf{cG1}, 
and this is discussed in appendix \ref{Large amplitude limit of the solution}.

\begin{table}[h!]
\setlength{\tabcolsep}{12pt}
\centering
\begin{tabular}{ c c c }
\hline \hline
$\cG_1$ & ${\rm sign} (\ch +\frac 14) $ & $a$, $b$ cond. 
\\ 
\hline \hline
\\
$\fr {1}{2 \pi R} \Big[ 
1 + \fr 12 \left( \fr { \ze_1 \ze_\al (1+2\ch) }{\sqrt{1+4\ch}}
-1 \right) e^{\pm i R/\la_{+}}$
& $+$ & $|\frac ab |<1 $ \\
$\pt{spacee}
- \fr 12 \left( \fr { \ze_1 \ze_\al (1+2 \ch )}{\sqrt{1+4\ch}}
+1 \right) e^{- R/\la_{-}} \Big]$
& &\\[20pt]
$\fr {1}{2 \pi R} \Big[ 
1 + \fr 12 \left( \fr { \ze_1 \ze_\al (1+2\ch) }{\sqrt{1+4\ch}}
-1 \right) e^{\pm i R/\la_{+}}$
& $+$ & $\frac ab > 1 $ \\
$\pt{spacee}
- \fr 12 \left( \fr { \ze_1 \ze_\al (1+2 \ch )}{\sqrt{1+4\ch}}
+1 \right) e^{\pm i R/\la_{-}} \Big]$
& &\\[20pt]
$\fr {1}{2 \pi R} \Big[ 
1 + \fr 12 \left( \fr { \ze_1 \ze_\al (1+2\ch) }{\sqrt{1+4\ch}}
-1 \right) e^{- R/\la_{+}}$
& $+$ & $\frac ab < -1 $ \\
$\pt{spacee}
- \fr 12 \left( \fr { \ze_1 \ze_\al (1+2 \ch )}{\sqrt{1+4\ch}}
+1 \right) e^{- R/\la_{-}} \Big]$
& &\\[20pt]
$\frac {1}{2 \pi R}  
- \frac {1}{2 \pi R} 
\exp \left( - \fr {R}{\la_1} \right)
\Bigg[ \cos \left( \frac {R}{\la_2} \right)
- \frac {\ze_1 (1+2 \ch)}{\sqrt{|1+4\ch|}} 
\sin \left( \frac {R}{\la_2} \right)
\Bigg]$
& $-$ & N/A \\
\\
\hline \hline
\end{tabular}
\caption{
Four cases of the general solution \rf{cG1}
for the Newtonian potential Green function $\cG_1$, 
categorized by conditions on the coefficient combinations $k_1$, $k_2$, and $k_3$.
Here $a=\ze_\al (k_1 +2 k_2)$ 
and $b=|k_1|\sqrt{1+4\ch}$.
The length scales are 
$\la_{\pm}=\sqrt{\fr {2|k_2^2-k_1 k_3|}{|a \pm b|} }$
and $\la_{1,2}=2\sqrt{
\fr {(k_2^2-k_1 k_3)}{(2 \sqrt{(k_2^2-k_1 k_3)}-k_1-2k_2)}}$.
For the last row, 
the case of $\ch+\frac 14 < 0$ implies
$k_2^2-k_1 k_3 >0$.
}
\label{solntable1}
\end{table}

To end this subsection, 
we revisit the equations \rf{SMEcoupled1} using a perturbative method adopted
in past works.
This method amounts to assuming the metric components can be obtained from a series
$h_\mn=h^{(0)}_\mn + h^{(1)}_\mn +... $.
We assume that the $0th$ order, GR solution, satisfies equations \rf{SMEcoupled1} for the case of vanishing $k_1$, $k_2$, and $k_3$.
Next we solve for the first correction to this solution
$h^{(1)}_\mn$.
Using this method, we find the zeroth and first correction for $h_{00}$ to be given by
\beq
h_{00} = 2U_0 + \ka (k_1+2k_2+k_3)\rho, 
\label{k1k2k3pert}
\eeq
where $U_0$ is the usual Newtonian potential from a mass density $\rho$ and
the first order correction is a contact term that is nonvanishing only within the mass distribution \cite{bkx15,Kostelecky:2021xhb}.
The first order solution \rf{k1k2k3pert} can be contrasted with the results of Table \ref{solntable1}.
Clearly the solutions in Table \ref{solntable1} represent a more detailed, careful look at the effects of the isotropic coefficients in \rf{constants}.

\section{Anisotropic exact solutions}
\label{anisotropic exact solutions}

While not the main focus of the paper, 
we discuss features of exact solutions
when the coefficients are anisotropic.
In the special limit that the only nonzero coefficients
in \rf{deM} are $k^{(6)0i0j0k0l}$, 
and still assuming the static matter situation,
the equations for $h_{jj}$
and $h_{00}$ can be decoupled.
In this case the equation satisfied by $h_{00}$
is given by
\beq
- \nabla^2 h_{00} 
-\frac 12 k^{(6)0i0j0k0l} \prt_{ijkl} h_{00} = 8\pi G_N \rh.
\label{h00aniso}
\eeq
An equation of this form was the subject of Refs.\ \cite{bkx15} and \cite{km17}, 
where a perturbative approach to the solution was taken
(with the result contained in \rf{potpert}).
In the perturbative approach, 
the second term in \rf{h00aniso}
is treated as much less than the first term, 
and a zeroth order solution is inserted in for $h_{00}$ in that term.
As one of the goals in this paper is to examine
the exact, 
nonperturbative solutions, 
we attempt here to look at anisotropic cases.

It turns out, 
exact analytic solutions for \rf{h00aniso} for the $15$ independent components of an arbitrary $k^{(6)0i0j0k0l}$
are quite challenging.
Instead we examine a special case
to show the features of such solutions.
We adopt a case where $k^{(6)0i0j0k0l}$
can be written in terms of a contraction 
$K^{ij}=k^{(6)0i0j0k0k}$ and its trace $K^{jj}$.
This reduces \rf{h00aniso} to the form,
\beq
- \nabla^2 ( 1 +\frac 37 K^{ij} \prt_{ij} - \frac {3}{70} K^{jj} \nabla^4 )
 h_{00} = 8\pi G_N \rh.
\label{h00aniso2}
\eeq

Next we assume only diagonal elements 
$K^{xx}$, $K^{yy}$, 
and $K^{zz}$, 
such that $K^{xx}=K^{yy}=K^{zz}/18$, 
then the equation can be written in the simpler form, 
\beq
\nabla^2  ( 1 - \lambda^2 \prt_z^2 ) h_{00} = -8\pi G_N \rh,
\label{zEqn}
\eeq
where $\lambda=\sqrt{8 |K_{zz}|/21}$
and $K_{zz}<0$, 
so one independent coefficient is left.
As before, 
we construct the Green function solution 
for a point source.
By writing the Green function version of equation \rf{zEqn} as two equations $( 1 - \lambda^2 \prt_z^2 ) \Ph = -\de^{(3)}(\vec r)$ and $\nabla^2 \cG=\Ph$, 
which can be solved separately \cite{arfken,Lindell},
and then combining the results,
we reduce the answer to an integral over one variable:
\beq
\cG (\rh, z ) = \fr {1}{8\pi \lambda}
\int^\infty_{-\infty} dz^\prime  \fr { e^{-|z^\prime|/\la }}{\sqrt{\rh^2 + (z-z^\prime)^2}},
\label{AnisoGrn}
\eeq
where we adopt cylindrical coordinates $\rh, \ph, z$
and $K_{zz}<0$.
If instead we consider the case of 
$K_{zz}>0$, 
there is a sign change in \rf{AnisoGrn} and
the integral changes to
\beq
\cG (\rh, z ) = \fr {-i}{8\pi \lambda}
\int^\infty_{-\infty} dz^\prime  \fr { e^{i|z^\prime|/\la }}{\sqrt{\rh^2 + (z-z^\prime)^2}}.
\label{AnisoGrn2}
\eeq
We have been unable to evaluate these integrals analytically, 
so we take a numerical approach.

\begin{figure}[h!]
\begin{center}
\includegraphics[width=4in]{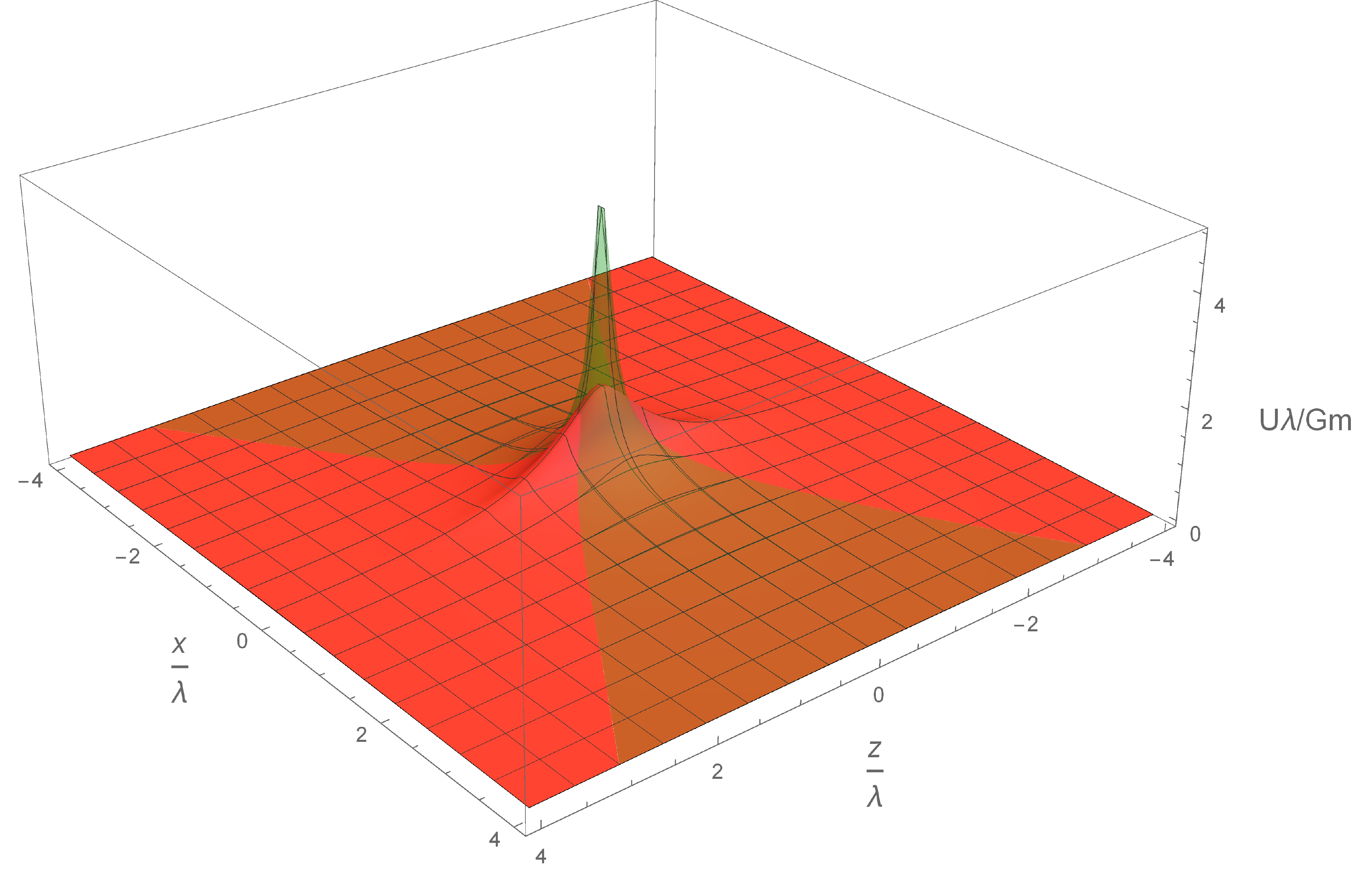}
\caption{\label{anisoplot1}Plot of the scaled anisotropic potential
of \rf{AnisoGrn} (damped case), shown in light red versus the Newtonian potential, in transparent green.
The source is a point mass at the origin.
The behavior is the same in the $y$ direction by the azimuthal symmetry of \rf{AnisoGrn}.}
\end{center}
\end{figure}

\begin{figure}[h!]
\begin{center}
\includegraphics[width=5in]{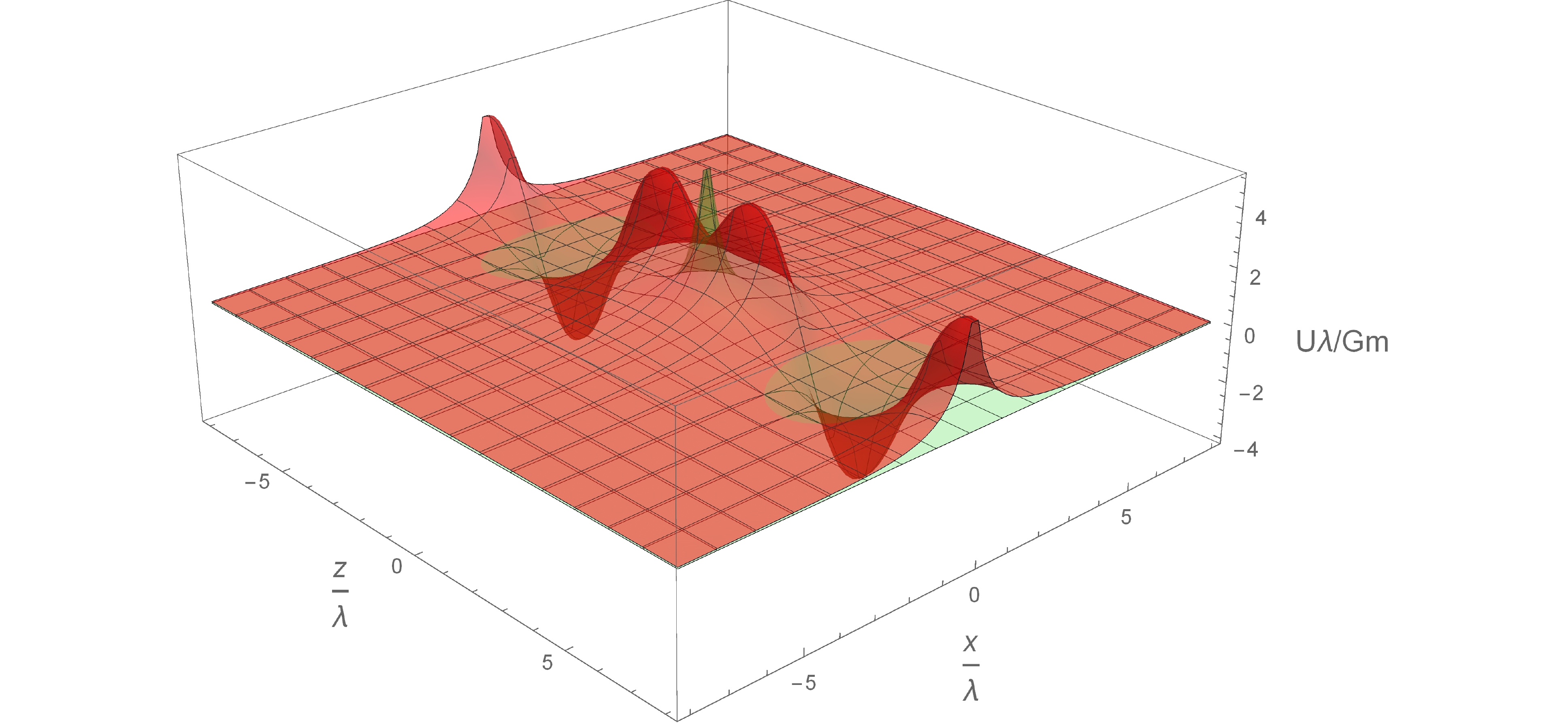}
\caption{\label{anisoplot2}Plot of the scaled anisotropic potential
of \rf{AnisoGrn2} (oscillating case), 
shown in light red versus the Newtonian potential, 
in transparent green.
The source is a point mass at the origin.
The behavior is the same in the $y$ direction by the aximuthal symmetry of \rf{AnisoGrn2}.}
\end{center}
\end{figure}

In the figures \ref{anisoplot1} and \ref{anisoplot2},
we plot the results from \rf{AnisoGrn} and \rf{AnisoGrn2}.
The standard Newtonian result $1/r$ is plotted
along with a numerical evaluation of \rf{AnisoGrn}
and \rf{AnisoGrn2}.
In the case of the damped-type solution in \rf{AnisoGrn}, 
we see a narrowed or cuspy behavior of the potential along the $x$ or $y$ direction, 
and the amplitude is reduced.
In the other case of the oscillating-type solution
in \rf{AnisoGrn2}, 
we see large oscillations along the $z$ direction 
that do not fall off rapidly.

To contrast with the numerically generated \rf{AnisoGrn} and \rf{AnisoGrn2}, 
we outline some features of the perturbative approach.
The idea is to solve for the Green function iteratively $\cG = \cG_{(0)} + \cG_{(1)} + \cG_{(2)} +... $, 
where the subscript indicates powers of $\la^2$.
The equations for the $0th$ and subsequent orders 
are given by
\bea
\nabla^2 \cG_{(0)} &=& -\de^{(3)}(\vec r),
\nonumber\\
\nabla^2 \cG_{(1)} &=& \la^2 \prt_z^2 \nabla^2 \cG_{(0)}, 
\nonumber\\
...
\nonumber\\
\nabla^2 \cG_{(n)} &=& \la^{2n} \prt_z^{2n} \nabla^2 \cG_{(n-1)}.
\label{pertscheme}
\eea
This type of approach is what led to the results in equation \rf{potpert}, 
where only the first order term is used but arbitrary coefficients assumed.

The $n^{th}$ term in the series \rf{pertscheme}
can be solved by using standard results for 
the derivatives of $1/r$ \cite{pw14}.
For $\cG_{(n)}$ we find
\beq
\cG_{(n)} = \fr {\la^{2n}}{4\pi} \fr {(4n-1)!!}{r^{2n+1}} n^{<zzz...>},
\label{nth}
\eeq
where $n^j=r^j/r$ is a unit vector and 
$n^{<jkl...>}$ is a symmetric trace free (STF) tensor formed from unit vectors and $\de^{ij}$
(for example, $n^{<jk>}=n^j n^k - (1/3) \de^{ij}$).
The STF tensor in \rf{nth} is to be evaluated along the $z$ direction.
Note that the convergence of such a series 
is not clear, 
since, 
for example,
the size of successive terms grow with $n$.

\begin{figure}[h!]
\begin{center}
\includegraphics[width=4in]{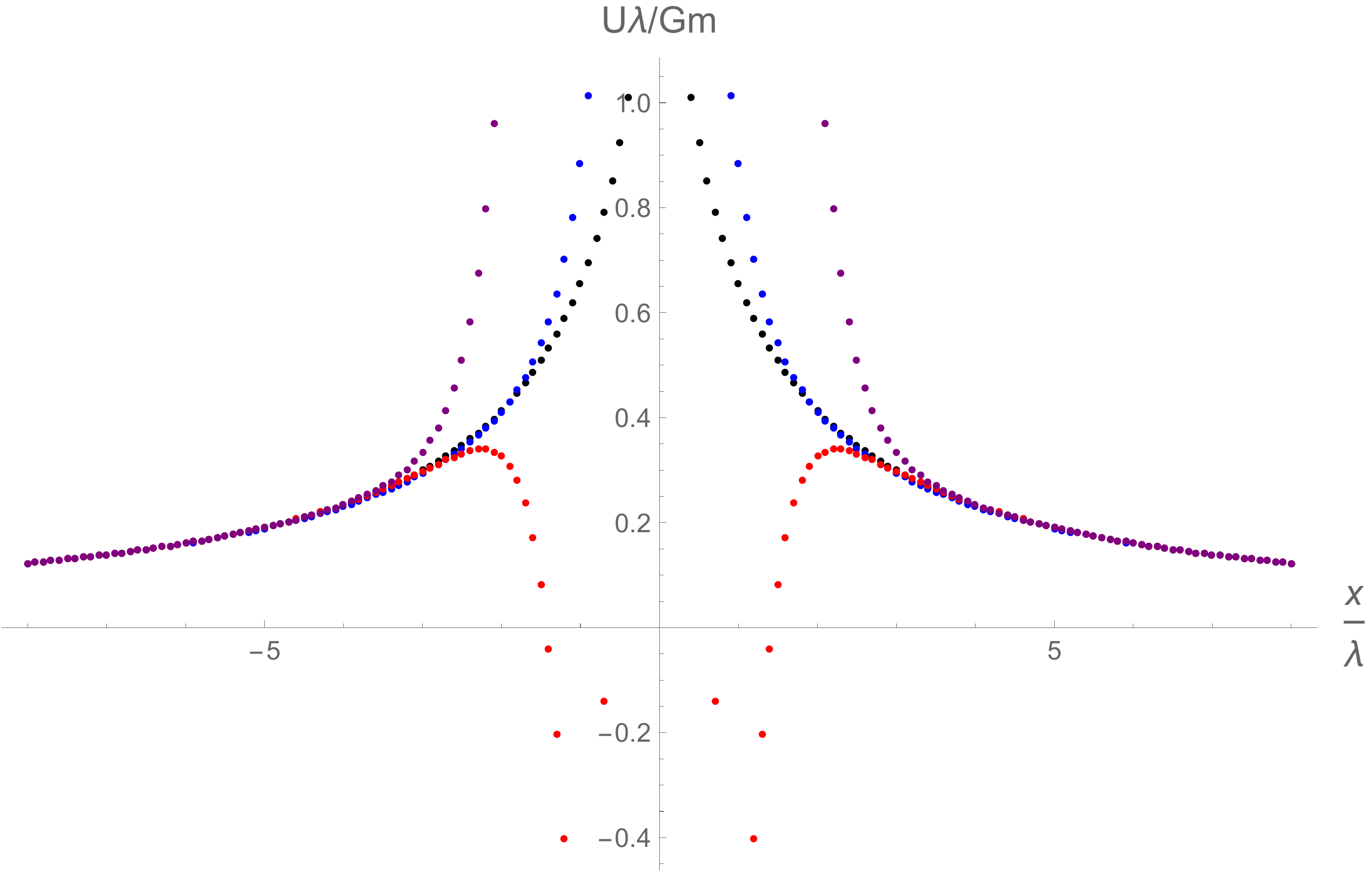}
\caption{\label{anisoapprox}Plot of the potential of \rf{AnisoGrn}
along with successive approximations in a perturbative scheme from \rf{pertscheme}.  
In black is the exact solution, 
blue is the first approximation
$\cG_0 + \cG_1$,
red is the second approximation $\cG_{(0)} + \cG_{(1)} +\cG_{(2)}$, 
and purple is $\cG_{(0)} + \cG_{(1)} +\cG_{(2)} +\cG_{(3)}$.}
\end{center}
\end{figure}

To illustrate this, 
we plot the exact numerical evaluation of \rf{AnisoGrn}
with the successive approximations \rf{pertscheme}. 
Figure \ref{anisoapprox} shows the approximations up to the third term in the series.
While the approximations approach the exact answer as $x/\la$ decreases, 
they vary considerably at scales of order $x \sim \la$.
In fact, 
it appears successive terms added to the first $G_{(1)}$
are worse than the just the first approximation alone!

From this brief study we can draw several conclusions
and open an area for future work.
We find that in the case of the damped exponential, 
where the equation to solve is 
$( 1 - \lambda^2 \prt_z^2 ) \nabla^2 \cG_1 = -\de^{(3)}(\vec r)$,
the first order approximation follows the exact solution until the $\rh$ and $z$ reach the scale of $\la$.
This behavior is expected and justifies
the use of the perturbative method generally.
On the other hand we see from Figure \ref{anisoapprox}, 
successive terms in a series \rf{pertscheme}
appear to fail to converge to the exact result.
It would be of interest to study in detail how well these approximations could follow an exact solution in general.

Of course, 
without knowing the true nature of the Newtonian level potential at short ranges from an unknown fundamental theory, 
we can only speculate.
Suppose, 
hypothetically,
that the potential in equations \rf{AnisoGrn} or \rf{AnisoGrn2} was indeed the potential coming from an underlying theory of physics.
The question then is how well a perturbative approach
could match this in the appropriate range.
We see above that for some choices of constants, 
the perturbative approach does not capture the behavior correctly.
However, 
there is an important caveat to include.
We truncated the expansion in \rf{general EFT case}
to mass dimension $6$. 
In the perturbative approach, 
beyond a first order approximation to the equation 
$( 1 + \lambda^2 \prt_z^2 ) \nabla^2 \cG_1 = -\de^{(3)}(\vec r)$, 
would necessitate the inclusion of mass dimension $8$
terms in the action, for consistency.
Indeed, 
it could be that higher order terms in the action could contribute to an approximation scheme, 
and provide ``counterterms" that result in 
smooth connection to the underlying potential
\cite{Kostelecky:2000mm}.
For example, 
imposing requirements term by term
in a series expansion, 
could place theoretical constraints
on the coefficients themselves.
It would be of interest to attempt
a general study of this in gravity
or other sectors like the photon sector.
Furthermore, 
this paper studies only the static limit, 
so it is of interest to study these issues
in the time-dependent limit.

Currently, 
experimental constraints on many of the anisotropic 
coefficients already exist using experiments
that satisfy the experimental constraint
of being sensitive enough to measure
the Newtonian forces between test masses
\cite{Yang:2012zzb}.
Thus if we assume that the perturbative 
approach is valid,
then the coefficient space for anisotropic 
coefficients is well covered in SR gravity tests \cite{Long:2014swa,Shao:2015gua,Shao:2016cjk,Shao:2018lsx}.

\section{Experimental implications}
\label{experimental implications}

\begin{figure}[h!]
\begin{center}
\includegraphics[width=3.5in]{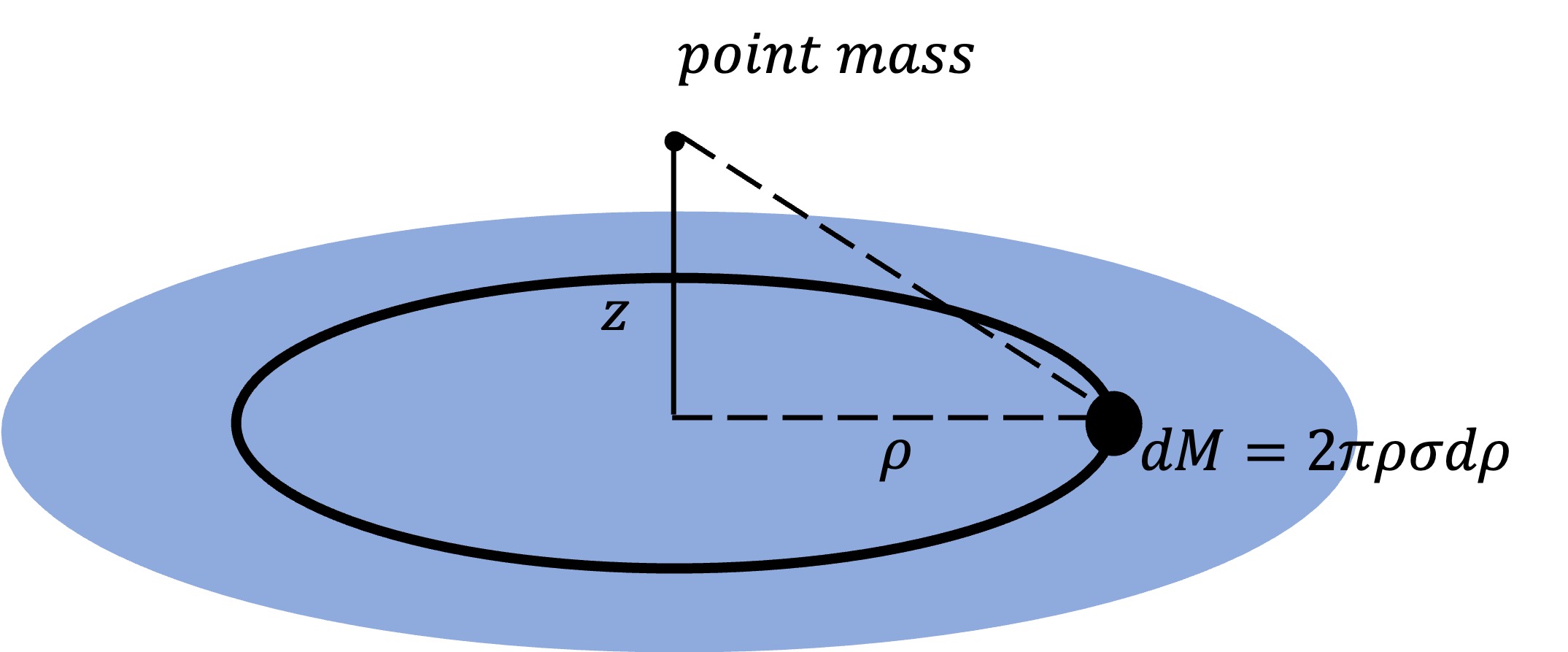}
\caption{\label{disk} A point mass lies above a thin disk of mass.
The gravitational field $\vec g = \vec \nabla U$ is integrated over this distribution.}
\end{center}
\end{figure}

\begin{figure}[h!]
\begin{center}
\includegraphics[width=5in]{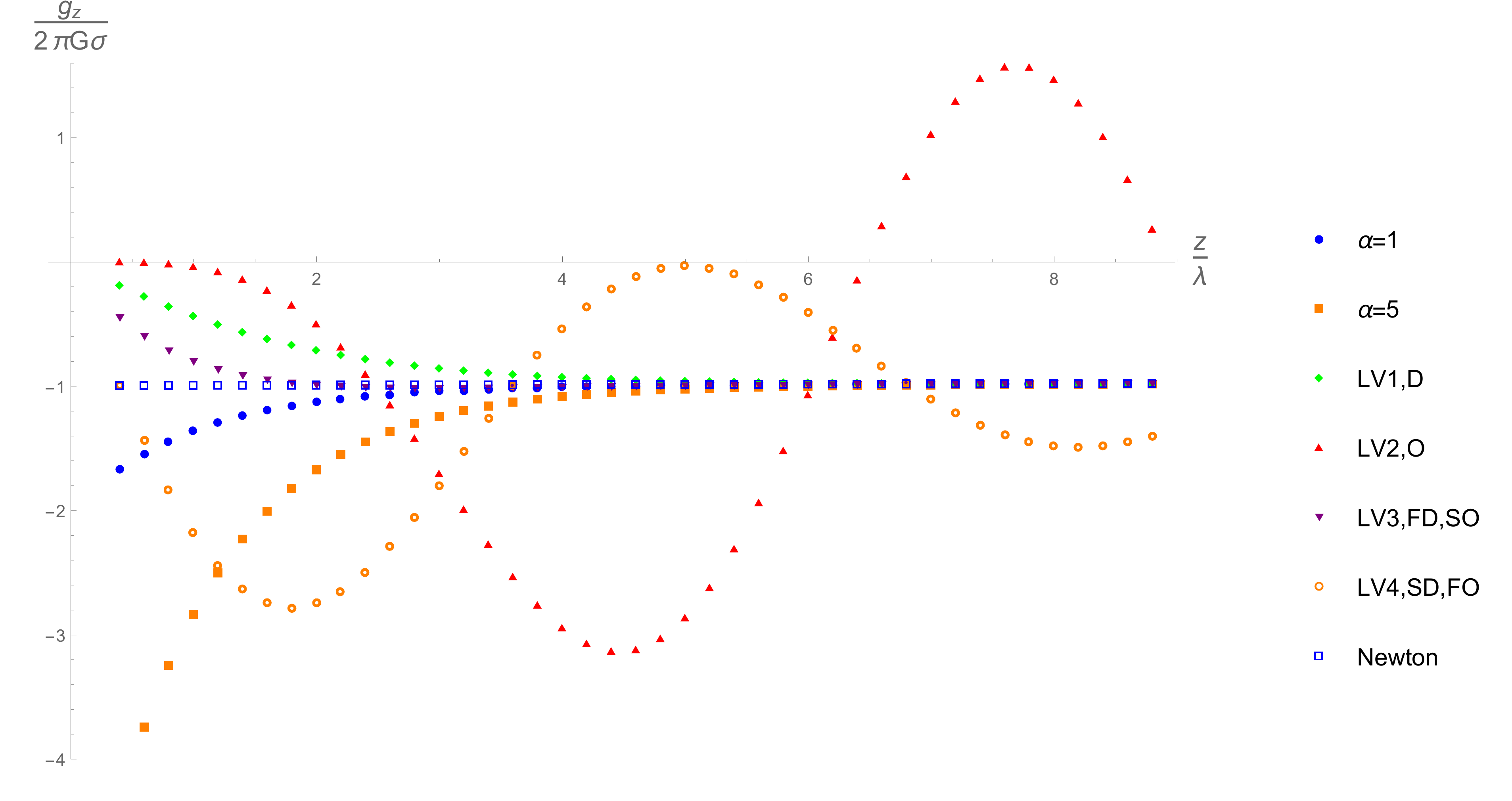}
\caption{\label{gplot} A plot of the gravitational field $g_z=\prt U/\prt z$
at a point above the center of a large flat disk of mass.  
We include two samples of the Yukawa force with $\al=1,5$ and the $4$ types of solutions in equations \rf{sample1} and \rf{sample2}.
The vertical axis is the scaled gravitational field and the horizontal axis is the height above the disk $z$ in units relative to the length scale $\la$.
We set the scale of Lorentz violation to be $\sqrt{|k_2|}=\la$.
The $4$ cases of Lorentz violation show
damping only (D), 
oscillation only (O), 
and fast damping (FD) with slow oscillation (SO) and finally slow damping (SD) with fast oscillation (FO).}
\end{center}
\end{figure}

Typical short range gravity tests 
are designed to 
measure the attraction 
between two masses, 
for instance two flat plates \cite{Long:2003dx,Yang:2012zzb}.
To see what implications the results of section 
\ref{isotropic Newtonian limit, exact solutions}
have on experimental signatures, 
we plot the gravitational field above a circular disk of mass (figure \ref{disk}).
We include the cases of the Newtonian gravity, 
the Yukawa potential term \rf{yukawa}, 
and the 4 sample cases of spacetime-symmetry breaking of equations \rf{sample1} and \rf{sample2} and display the vertical component in figure \ref{gplot}.

\begin{figure}[h!]
\begin{center}
\includegraphics[width=5in]{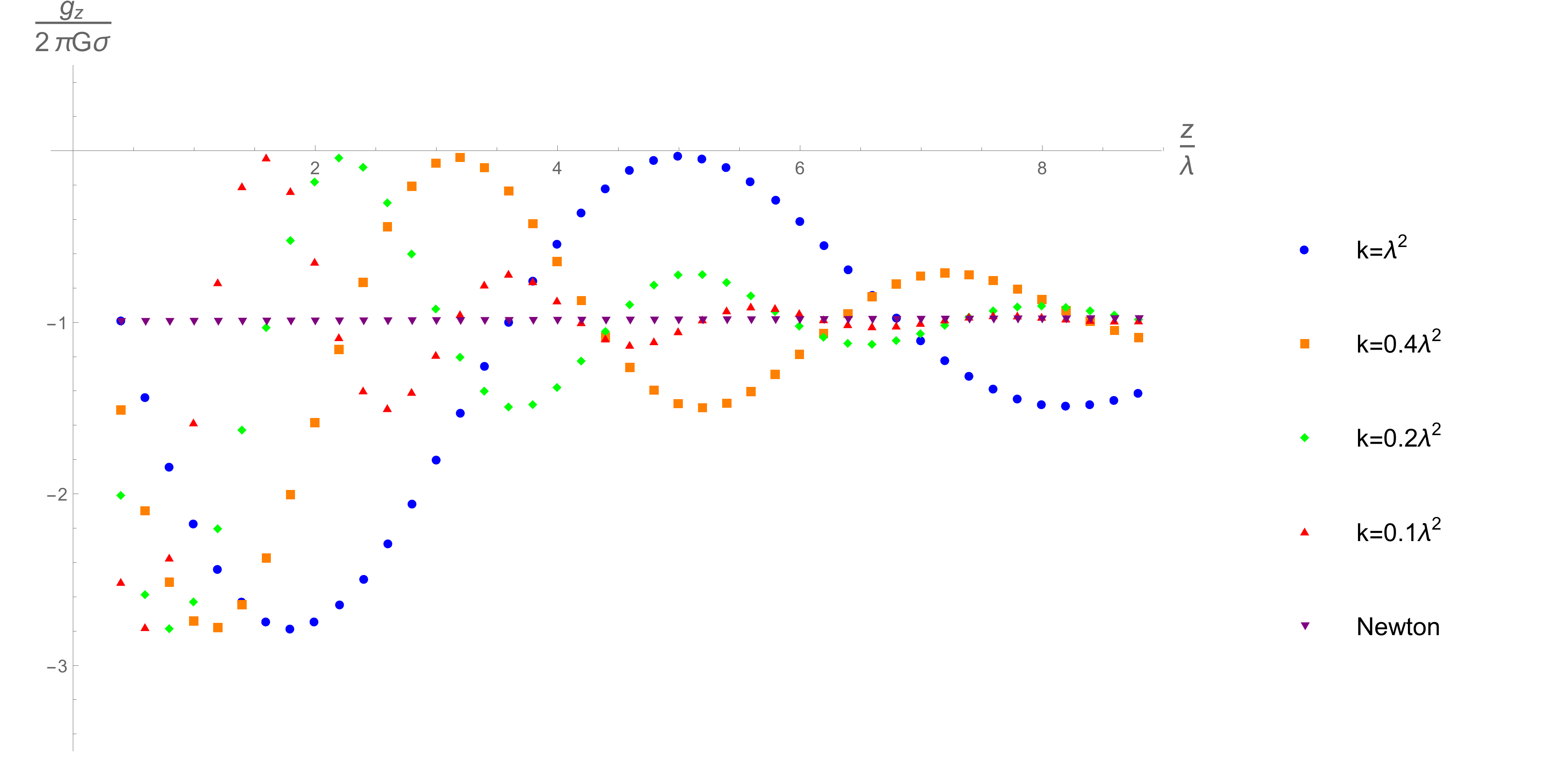}
\caption{\label{gplot2} A plot of the gravitational field $g_z=\prt U/\prt z$
at a point near a large flat plate of mass.  
In this plot we let $k_2$ vary and show the effect on the damped and oscillating solution from equation \rf{sample2}.}
\end{center}
\end{figure}

It is clear that the cases studied here exhibit behavior quite different from the Yukawa parametrization.
The Yukawa parametrization shows a deviation from the Newtonian case with the force becoming stronger on shorter scales, 
as expected.
The different Lorentz violation cases have oscillatory behavior with and without damping.
To get an idea how analysis might proceed, 
we produce the same plot with one of the $4$ cases, the damped and oscillating solution, 
but with varying values of $|k_2|$, 
in plot \ref{gplot2}.
As $|k_2|$ becomes smaller, 
the effects deviating from the Newton case narrow to a region at smaller and smaller length scales.
This shows that some of the solutions have the feature that $|k_2|$ could be constrained to be below a certain length scale.
For example, 
we can make a crude estimate
from Figures \ref{gplot} and \ref{gplot2}:
if the Yukawa-type force has been constrained to a region of standard $\al-\la$ space where $\al \sim 1$ and $\la \sim 200 \mu m$, 
like the experiment in reference \cite{Lee20}, 
than roughly $\sqrt{k_2} < 200 \mu m$, 
if one used the specific case of \rf{sample1}.

However, 
what we plot here in Figure \ref{gplot2} is only a one coefficient special case, 
in the full model one could use each of the cases in Table \ref{solntable1} to fit data and rule out a region of $k_1-k_2-k_3$ space,
similar to the way exclusion regions are mapped out in $\al-\la$ plots in the experimental literature.
In general, the region of $k_1-k_2-k_3$ affects
the amplitude of the exponential and oscillatory terms 
and the length scales involved, 
as can be seen from Table \ref{solntable1}.

\section{Summary and outlook}
\label{summary and outlook}

In this paper, 
we studied short-range gravity signals
for Lorentz violation that go beyond 
the leading order approximation,
by taking a non-perturbative approach.
The focus was on isotropic coefficients, 
since they are generally harder 
to measure in experiments and observations.
The main results are the coupled field
equations for the metric components $h_{00}$ and $h_{jj}$ in the static limit, 
equations \rf{SMEcoupled1}, 
and the general solution for the Green function $\cG_1$ for $h_{00}=2U$, 
organized into four cases in Table \ref{solntable1}.

These solutions go beyond the standard Yukawa parametrization \rf{yukawa} and could be studied in short-range gravity tests of all kinds.
Particularly, 
it may be of interest for short-range tests 
that probe large non-Newtonian forces.
One option for data analysis is to restrict attention to the solutions in the last two rows in Table \ref{solntable1}, 
which do not exhibit undamped oscillations.
One could then attempt to use experimental data to measure the coefficient combinations $k_1$, $k_2$, and $k_3$ (equation \rf{constants}) contained in these expressions. 
Analysis could also proceed with a simpler two coefficient special case model in section \ref{Special case model}, 
or, 
in the case of tests sensitive to very large non-Newtonian forces, 
one could use the large amplitude, 
$\ch=-1/4 + \ep$ limit,
outlined in the appendix and Table \ref{solntable2}.

The isotropic mass dimension $6$ 
coefficients that can be probed using the solutions in this work, 
in equations \rf{constants},
appear to be distinct combinations
from the combination appearing in gravitational wave propagation tests \cite{Mewes:2019}, 
as shown in \rf{k00}.
This demonstrates the usefulness of 
additional short-range gravity test analysis 
outlined in this work, 
providing an independent probe 
of isotropic coefficients from GW tests.

In this work, 
we also collect some useful pedagogical results with explicit examples of the construction of
isotropic coefficients from Young Tableau, 
as discussed in appendix \ref{isotropic limit of coefficients}.
In section \ref{anisotropic exact solutions} we discussed exact solutions in the case of anisotropic coefficients, 
and compared the results to perturbative
methods used so far.
It would be of interest to compute integrals like \rf{AnisoGrn} and \rf{AnisoGrn2} analytically, 
if possible.
In addition, 
a study of the convergence of the series \rf{pertscheme} and related topics like 
adding time dependence would be of interest.
Considerations of this paper 
could be applied to the photon sector \cite{km09,bk04}, 
where, 
analogous to gravity, 
new types of massive photon-like signals may be revealed.

\section{Acknowledgements}

The authors acknowledge support from the National Science Foundation under grants 1806871 and 2207734.
We thank Embry-Riddle Aeronautical University's Undergraduate Research Institute for support of Jennifer L.\ James and Janessa R.\ Slone.
Valuable comments on the manuscript have been provided by Brett Altschul, V.\ Alan Kosteleck\'y, 
and Rui Xu.



\section{Appendix}

\subsection{Special Case Model}

We record the exact field equations
for the action in \rf{action1},
$\cL = \frac {1}{2\ka} \sqrt{-g} \left( R + k_\ab R^\ab R \right)$.
Upon variation with respect to the full metric $g_\mn$
we obtain:
\bea
G^\mn &=& \frac 12 g^\mn k_\ab R^\ab R 
- k_\al^{\pt{\al}\mu} R^{\nu\al} R 
- k_\al^{\pt{\al}\nu} R^{\mu\al} R 
- k_\ab R^\ab R^\mn
\nonumber\\
&&
-\frac 12 g^\mn \nabla^\al \nabla^\be (k_\ab R) 
-g^\mn \nabla^2 ( k_\ab R^\ab )
\nonumber\\
&&
+\frac 12 \nabla_\al \nabla^\mu ( k^{\nu \al} R )
+\frac 12 \nabla_\al \nabla^\nu ( k^{\mu \al} R )
\nonumber\\
&&
+ \nabla^\mu \nabla^\nu (k_\ab R^\ab )
-\frac 12 \nabla^\al \nabla_\al ( k^\mn R ).
\label{k2case}
\eea
Here we treat $k_{\al\be}$ as a fixed background set of coefficients
and do not consider field equations obtained with the variation $\de k_{\al\be}$, 
but this could be generalized.

\subsection{Differential equation results}

Here we record some basic results that we use 
in constructing the general solutions 
for the PDE's in the paper.
Boundary conditions are assumed where
the fields vanish at spatial infinity.
First we note the Helmholtz equation for a field $\ps$
\beq
    (\nabla^{2}+\om^2)\ps = -\de^{(3)} (\vec R).
    \label{helmholtz}
\eeq
This is solved with the following Green function (e.g., see Ref.\ \cite{arfken})
\beq
    \ps = \fr {e^{\pm i\om R}}{4\pi R}.
    \label{Helmholtzsoln}
\eeq
Note that if $\om$ is a general complex number, 
$\om=a+ib$ then one obtains
\beq
    \ps = \fr {e^{ \pm iaR}}{4\pi R} e^{\mp b R},
    \label{Helmholtzsoln2}
\eeq
which shows that oscillation and damping or growth can occur.
When $a=0$ and $\mp b <0$, then the solution to the Proca or modified Helmholtz equation is recovered.

We also record here the less common nonlocal equation considered long ago in generalized electrodynamics \cite{Podolsky:1942zz,Pais:1950za},
\beq
    (\nabla^2 - \la^2 \nabla^4 )\ps = - \de^{(3)} (\vec R)
    \label{nl1}.
\eeq
For this equation, 
the Green function is (e.g., see Ref.\ \cite{Lindell}),
\beq
    \psi = \fr{1}{4\pi R} - \fr{e^{-R/\la}}{4\pi R}.
    \label{nlsoln}
\eeq
Here again, 
$\la$ could be complex, 
yielding oscillatory behavior.

\subsection{Isotropic limit of coefficients}
\label{isotropic limit of coefficients}

We record here the portion of the field equations in the static limit
involving the $\de M^{\mu\nu\rh\si} h_{\rh\si}$ (definition in Eq.\ \rf{deM}).
It is useful in what follows to enumerate the specific tensor symmetries of the coefficients involved \rf{sqk}.
For convenience, 
we display here the young tableau \cite{km16} 
for the coefficients with spacetime indices:
\bea
s^{(4)\mu\rh\ep\nu\si\ze} &\Leftrightarrow& \young(\mu \nu,\rh \si,\ep \ze), \nonumber\\[10pt]
q^{(5)\mu\nu\ka\al\be\ga\de} &\Leftrightarrow &  
\young(\mu\al\ga,\nu\be\de,\ka) ,
\nonumber\\[10pt]
s^{(6)\mu\rh\ep\nu\si\ze\al\be} &\Leftrightarrow& 
\young(\mu\nu\al\be,\rh\si,\ep\ze),
\nonumber\\[10pt]
k^{(6)\mu\ep\nu\ze\rh\al\si\be} &\Leftrightarrow& 
\young(\mu\nu\rh\si,\ep\ze\al\be).
\label{spacetimeyoung}
\eea
Later below, 
we break down these coefficients into spatial subsets.
Young tableau and the process of breaking down Tableau into representations of subgroups is described elsewhere \cite{lichtenberg,hamermesh}.

For the space and time components $\de M^{00\rh\si} h_{\rh\si}$, $\de M^{0i\rh\si} h_{\rh\si}$, and $\de M^{ij\rh\si} h_{\rh\si}$, 
we obtain
\bea
\de M^{00\rh\si}h_{\rh\si} &=& 
-\frac 12 k^{(6)0i0j0k0l} \prt_{ijkl} h_{00}
\nonumber\\
&&
-\{
\frac 12 k^{(6)0j0k0lim} \prt_{jklm} 
+ \frac 14 q^{(5)0ij0k0l} \prt_{jkl} 
\} 
h_{0i}
\nonumber\\
&&
-\{ 
\frac 12 [
s^{(4)0ik0jl} \prt_{kl} + s^{(6)0ik0jlmn} \prt_{klmn} 
+k^{(6)0k0limjn} \prt_{klmn} ]
\nonumber\\
&&
\pt{-- }+ \frac 14 [q^{(5)0ik0ljm} + q^{(5)0jk0lim}] \prt_{klm} 
\}
h_{ij},\\[10pt]
\de M^{0i\rh\si}h_{\rh\si} &=& 
-\{ \frac 12 k^{(6)0kil0m0n} \prt_{klmn} 
-\frac 14 q^{(5)0ik0l0m} \prt_{klm} \}
h_{00}
\nonumber\\
&&
\hspace{-20pt}
+\{ \frac 14 [ s^{(4)0ik0jl} \prt_{kl} + s^{(6)0ik0jlmn} \prt_{klmn}]
-\frac 12 k^{(6)0kilomjn} \prt_{klmn}
\nonumber\\
&&
-\frac 18 [-q^{(5)0ik0ljm} 
+ q^{(5)0jk0lim}
+q^{(5)ijk0l0m}] \prt_{klm} \} h_{0j}
\nonumber\\
&&
\hspace{-20pt} 
-\{ \frac 12 [ s^{(4)0jlikm} \prt_{lm} 
+ s^{(6)0jlikmnp} \prt_{lmnp}]
+\frac 12 k^{(6)0limjnkp} \prt_{lmnp} 
\nonumber\\
&&
+\frac 14 [ q^{(5)0jlimkn} 
+ q^{(5)ijl0mkn}] \prt_{lmn} \} h_{jk},\\[10pt]
\de M^{ij\rh\si}h_{\rh\si} &=& 
-\{ \frac 12 [ s^{(4)0ik0jl} \prt_{kl} + s^{(6)0ik0jlmn} \prt_{klmn}]
\nonumber\\
&&
\hspace{-10pt}
+\frac 12 k^{(6)ikjl0m0n} \prt_{klmn}
+\frac 14 [ q^{(5)0ik0ljm}+ q^{(5)0jk0lim} ] 
\prt_{klm} \}
h_{00}
\nonumber\\
&&
-\{ \frac 14 [ s^{(4)i0ljkm} \prt_{lm} + s^{(6)i0ljkmnp} \prt_{lmnp}] +\frac 14 [ i \rightleftharpoons j] 
\nonumber\\
&&
\pt{++}
+\frac 12 k^{(6)0lkminjp} \prt_{lmnp}
+\frac 18 [ q^{(5)i0ljmkn} + q^{(5)j0limkn}
\nonumber\\
&&
\pt{++}
+q^{(5)ikljm0n} +q^{(5)jklim0n}] 
\prt_{lmn} \}h_{0k}
\nonumber\\
&&
\hspace{-20pt}
-\{ \frac 12 [ s^{(4)ikmjln} \prt_{mn} + s^{(6)ikmjlnpq} \prt_{mnpq}] 
+\frac 12 k^{(6)imjnkplq} \prt_{mnpq} 
\nonumber\\
&&
+\frac 14 [q^{(5)ikmjnlp} + q^{(5)jkminlp}] \prt_{mnp}
\}h_{kl},
\label{deMs_1}
\eea
where we have already used the tensor symmetry properties of the coefficients in \rf{sqk}.
To simplify the terms occuring in \rf{deMs_1}, 
we assume only isotropic coefficients and express each set of coefficients occurring in \rf{deMs_1} in terms of its 
scalar contractions.
To elucidate the process, 
the results for each of the coefficients are recorded in Tables \ref{isotable1} and \ref{isotable2} below.
Note that isotropic limits of the coefficients are of interest independently of the present paper.
This is due to the challenges of their measurement with the same precision as anisotropic coefficients 
\cite{Altschul:2006ka}.

\begin{table}[h!]
\setlength{\tabcolsep}{12pt}
\centering
\begin{tabular}{ c c c }
\hline \hline
Coefficients & Tableau & isotropic form 
\\ 
\hline \hline
\\
$s^{(4)0ik0jl}$ 
& $\young(ij,kl)$
& $\frac 16 (\de^{ij}\de^{kl}- \de^{kj} \de^{il}) s^{(4)0mn0mn}$ \\[5pt]
\hline
\\
$s^{(4)0jlikm}$
& $\young(ij,kl,m)$
& $0$ \\[5pt]
\hline
\\
$s^{(4)ikmjln}$ 
& $\young(ij,kl,mn)$
& $\frac 16 \ep^{ikm}\ep^{jln}s^{(4)pqrpqr}$
\\[5pt]
\hline
\\
$q^{(5)0ij0k0l}$ 
& $\young(ikl,j)$
& $0$ \\[5pt]
\hline\\
$q^{(5)0ik0ljm}$ 
& $\young(ijl,km)$
& $0$ \\[5pt]
\hline\\
$q^{(5)ijk0l0m}$
& $\young(ilm,j,k)$
& $-\fr 13 \ep^{ijk} \de^{lm} (^*q_0^{\pt{0}0n0n})$ \\[5pt]
\hline\\
$q^{(5)0jlimkn}$
& $\young(jik,lmn)$
& $0$ \\[5pt]
\hline\\
$q^{(5)ijl0mkn}$
& $\young(ikm,jn,l)$
& $0$ \\[5pt]
\hline\\
$q^{(5)ikmjnlp}$
& $\young(ijl,knp,m)$
& $-\fr 16 \ep^{ikm} 
(\de^{jl} \de^{np} -\de^{jp} \de^{nl} )
(^*q_0^{\pt{0}qrqr})$ \\[5pt]
\hline \hline
\end{tabular}
\caption{Mass dimension 4 and 5 coefficients 
expressed assuming isotropic coefficients only.  
The dual is defined by $^*q_\la^{\pt{\la}\al\be\ga\de}=
-\frac {1}{3!} \ep_{\la\mu\nu\ka} q^{\mu\nu\ka\al\be\ga\de}$ 
with $\ep_{0123}=+1$.}
\label{isotable1}
\end{table}

\begin{table}[h!]
\setlength{\tabcolsep}{12pt}
\centering
\begin{tabular}{ c c c }
\hline \hline
Coefficients & Tableau & isotropic form 
\\ 
\hline \hline
\\
$s^{(6)0ik0jlmn}$ 
& $\young(ijmn,kl)$
& $\frac {1}{60} 
( \de^{in} \de^{jm} \de^{kl}
+\de^{im} \de^{jn} \de^{kl}
-\de^{il} \de^{jn} \de^{km}
-\de^{il} \de^{jm} \de^{kn}
$\\
& &
$-\de^{in} \de^{jk} \de^{lm}
+\de^{ij} \de^{kn} \de^{lm}
-\de^{im} \de^{jk} \de^{ln}
+\de^{ij} \de^{km} \de^{ln}$\\
& & 
$-2 \de^{il} \de^{jk} \de^{mn}
+2 \de^{ij} \de^{kl} \de^{mn})
s^{(6)0pq0pqrr}
$ \\[5pt]
\hline
\\
$s^{(6)0jlikmnp}$
& $\young(ijnp,kl,m)$
& $0$ \\[5pt]
\hline
\\
$s^{(6)ikmjlnpq}$
& $\young(ijpq,kl,mn)$
& $\frac {1}{90} (
-\de^{iq} \de^{jp} \de^{kn} \de^{lm}
-\de^{ip} \de^{jq} \de^{kn} \de^{lm}$ \\
& & 
$+\de^{in} \de^{jq} \de^{kp} \de^{lm}
+...) s^{(6)rstrstuu}$ \\[5pt]
\hline
\\
$k^{(6)0i0j0k0l}$ 
& $\young(ijkl)$
& $\frac {1}{15} (\de^{ij} \de^{kl}
+\de^{ik} \de^{jl}+\de^{il} \de^{jk}) k^{(6)0m0m0n0n}$ \\[5pt]
\hline
\\
$k^{(6)0j0k0lim}$
& $\young(iljk,m)$
& 0 \\[5pt]
\hline\\
$k^{(6)0k0limjn}$
& $\young(ijkl,mn)$
& $\frac {1}{60} (
-2 \de^{in} \de^{jm} \de^{kl}
-\de^{in} \de^{jl} \de^{km}
-\de^{il} \de^{jm} \de^{kn}
-\de^{in} \de^{jk} \de^{lm}$\\
& &
$+\de^{ij} \de^{kn} \de^{lm}
-\de^{ik} \de^{jm} \de^{ln}
+\de^{ij} \de^{km} \de^{ln}
+\de^{il} \de^{jk} \de^{mn}$\\
& &
$+\de^{ik} \de^{jl} \de^{mn}
+2\de^{ij} \de^{kl} \de^{mn}
) k^{(6)0p0pqrqr}$ \\[5pt]
\hline\\
$k^{(6)0kil0mjn}$
& 
& $k^{(6)0kil0mjn} =k^{(6)0k0limjn} - k^{(6)0k0ilmjn}$\\[5pt] 
\hline\\
$k^{(6)0limjnkp}$
& \young(jkil,npm)
& $0$ \\[5pt]
\hline\\
$k^{(6)imjnkplq}$
& \young(ijkl,mnpq)
& $\frac {1}{240}
( 
2 \de^{iq}\de^{jp} \de^{kn} \de^{lm}
+ \de^{ip}\de^{jq} \de^{kn} \de^{lm}$\\
& &
$+ \de^{in}\de^{jp} \de^{kq} \de^{lm}
+...) k^{(6)rsrstutu} $ \\[5pt]
\hline \hline
\end{tabular}
\caption{Mass dimension 6 coefficients expressed assuming isotropic coefficients only.  For two of the sets of coefficients,
$s^{(6)ikmjlnpq}$ and $k^{(6)imjnkplq}$, 
the expression in terms of 
Kronecker deltas is abbreviated 
due to its length.
It can be calculated using the ``YoungProject" command of the {\it xTras} package for xTensor \cite{Nutma:2013zea}.}
\label{isotable2}
\end{table}

The results from the Tables \ref{isotable1}
and \ref{isotable2} are then inserted in the expressions \rf{deMs_1} and simplified to the following:
\bea
\de M^{00\rh\si}h_{\rh\si} &=& 
- \frac {1}{10} k^{(6)0m0m0n0n} \nabla^4 h_{00}
\nonumber\\
&&
+\{ 
\frac {1}{12} s^{(4)0kl0kl} 
+\frac {1}{30} s^{(6)0kl0klmm} \nabla^2  
+\frac {1}{30} k^{(6)0k0klmlm} \nabla^2 
\} 
\nonumber\\
&&
\times
(\prt_i \prt_j - \de_{ij} \nabla^2 ) h_{ij},
\\[10pt]
\de M^{0i\rh\si}h_{\rh\si} &=&
\{ 
\frac {1}{24} s^{(4)0kl0kl} 
+\frac {1}{60} s^{(6)0kl0klmm} \nabla^2  
-\frac {1}{30} k^{(6)0k0klmlm} \nabla^2 
\} 
\nonumber\\
&&
\times 
(\de_{ij} \nabla^2 -\prt_i \prt_j ) h_{0j}
\nonumber\\
&&
-\frac {1}{24}( ^*q_0^{\pt{0}0l0l}) \ep_{ijk} \nabla^2 \prt_j h_{0k},\\
\de M^{ij\rh\si}h_{\rh\si} &=&
\{ 
-\frac {1}{12} (s^{(4)0kl0kl} + \frac 12 s^{(4)klmklm})
\nonumber\\
&&
\pt{s}
-\frac {1}{30} (s^{(6)0kl0klmm} + k^{(6)0k0klmlm}) \nabla^2 
\nonumber\\
&&
\pt{s}
-\frac {1}{72} s^{(6)klmklmnn} \nabla^2
-\frac {1}{240} k^{(6)klklmnmn} \nabla^2
\} \de^{ij} \nabla^2 h_{00}
\nonumber\\
&&
+\{ 
\frac {1}{12} (s^{(4)0kl0kl} + s^{(4)klmklm})
\nonumber\\
&&
\pt{s}
+\frac {1}{30} (s^{(6)0kl0klmm} + k^{(6)0k0klmlm}) \nabla^2 
\nonumber\\
&&
\pt{sm}
+\frac {1}{36} s^{(6)klmklmnn} \nabla^2
-\frac {1}{240} k^{(6)klklmnmn} \nabla^2
\} \prt_{ij} h_{00}
\nonumber\\
&&
-\{ 
\frac {1}{24} s^{(4)klmklm}
+\frac {1}{72} s^{(6)klmklmnn} \nabla^2
\nonumber\\
&&
\pt{sm}
+\frac {1}{240} k^{(6)klklmnmn} \nabla^2
\} \de^{ij} \nabla^2 h_{pp}
\nonumber\\
&&
+\{ 
\frac {1}{12} s^{(4)klmklm})
+\frac {1}{36} s^{(6)klmklmnn} \nabla^2
\nonumber\\
&&
\pt{sm}
-\frac {1}{60} k^{(6)klklmnmn} \nabla^2
\} \nabla^2 h_{ij}
\nonumber\\
&&
+ \frac {1}{80} k^{(6)klklmnmn} \nabla^2 \prt_{ij} h_{pp}
\nonumber\\
&&
-\frac {1}{24} ^*q_0^{\pt{0}lnln} 
(\ep_{mki} \prt_m \nabla^2 h_{jk} +\ep_{mkj} \prt_m \nabla^2 h_{ik}).
\label{deMs_2}
\eea
Note that for the last of 
equations \rf{deMs_2}, 
one takes the trace in $ij$ to obtain 
the result \rf{SMEcoupled1}.
The equations for the off-diagonal components 
$h_{ij} - \frac 13 \de_{ij} h_{kk}$ can be obtained from \rf{deMs_2} by subtracting the trace appropriately.
Like the sample case in Section \ref{Special case model} 
(see equation \rf{offdiag})
$h_{ij} - \frac 13 \de_{ij} h_{kk}$ is sourced by $h_{00}$
and $h_{jj}$ and since $h_{00}$ is of primary interest in this work
we do not include the solution here.

To compare the results with those obtained
by looking at propagation effects in gravitational waves, 
we record here the isotropic combination
of coefficients in the gravity sector called
$(k_{I}^{(6)})_{00}$.
The spherical coefficients are  defined by 
(see reference \cite{Mewes:2019}),
\beq
\frac 12 ( {\hat s}^{+-+-} + {\hat k}^{++--} )
= \sum _{jm} \om^4 (-1)^j Y_{jm} (\hat v) (k_{I}^{(6)})_{jm},
\label{k6iso}
\eeq
where the coefficients on the left hand side are
to be evaluated at $d=6$ only, 
and from \rf{sqk} with the substitution $\prt_\mu \rightarrow -i p_\mu $.
The $+$, $-$, and the $\hat v$ refer to a helicity basis for
GW's so that $p_\mu = (-1, \hat v)$.
A plus or minus in an index indicates a contraction with 
the helicity basis vectors ${\bf e}^{+}$ and ${\bf e}^{-}$ (see Ref.\ \cite{Mewes:2019, km09} for more details).
Focusing on only the isotropic piece $(k_{I}^{(6)})_{00}$
it can be shown the following relation holds:
\bea
(k_{I}^{(6)})_{00} &=& \sqrt{4\pi} \Big[ 
-\frac {1}{12} ( s^{(6)0ij0ij00} + s^{(6)ijkijk00} 
+\frac 15 s^{(6)0ij0ijkk} 
+ \frac 13 s^{(6)ijkijkll} )
\nonumber\\
&&
\pt{space}
+\frac {1}{60} ( 4 k^{(6)0i0i0j0j} + 8 k^{(6)0i0ijkjk} 
+k^{(6)ijijklkl} ) \Big].
\label{k00}
\eea

\subsection{Large amplitude limit of the solution}
\label{Large amplitude limit of the solution}

Given the results of the section 
\ref{general EFT case}, 
we record here the large amplitude limit, 
where $\ch \sim -1/4$.
Specifically we let $\ch  = -1/4 +\ep$
and explore the solutions for small $\ep$.
When simplifying the solution \rf{cG1} in this limit, 
the result depends on the sign of 
the coefficient combination $k_2+2k_3$ 
and the sign of $\ep$.
Thus the result breaks into 4 cases.
Specifically, 
when expanding to the lowest order in $\ep$
we find the four solutions in the Table \ref{solntable2}.

\begin{table}[h!]
\setlength{\tabcolsep}{12pt}
\centering
\begin{tabular}{ c c c }
\hline \hline
$\cG_1$ & ${\rm sgn} (k_2 +2 k_3 )$  & ${\rm sgn} \ep$  \\ 
\hline \hline
$\frac {1}{2 \pi R} 
\Bigg[ 1 - \fr 12 
\exp \left( -\frac{R}{\sqrt{|k_2+2k_3|}} \right)
\Big[ 2 + \ps \frac {R}{\sqrt{|k_2+2k_3|}} \Big] $
& $+$ & $+$ \\[20pt]
$\frac {1}{2 \pi R} 
\Bigg[ 1 - \fr 12 
\exp \left( \frac{\pm i R}{\sqrt{|k_2+2k_3|}} \right)
\Big[ 2 \pm i \ps \frac {R}{\sqrt{|k_2+2k_3|}} \Big] $
& $-$ & $+$ \\[20pt]
$\frac {1}{2 \pi R}  
- \frac {1}{2 \pi R} 
\exp \left( - \fr {R}{\sqrt{|k_2+2k_3|}} 
\right)
\Bigg[ \cos \left( 2 \ps \sqrt{\fr {|\ep|}{|k_2+2k_3|}} R \right)$
& $+$ & $-$ \\
$\pt{space space}
+ \frac {1}{4\sqrt{|\ep|}} 
\sin \left( 2 \ps \sqrt{\fr {|\ep|}{|k_2+2k_3|} } R \right)
\Bigg]$
& & \\ [20pt]
$\frac {1}{2 \pi R}  
- \frac {1}{2 \pi R} 
\exp \left( 
- 2 \ps \sqrt{ \fr {|\ep|}{|k_2+2k_3|} } R 
\right)
\Bigg[ \cos 
\left( \fr {R}{\sqrt{|k_2+2k_3|}} 
\right)$
& $-$ & $-$ \\
$\pt{space space}
+ \frac {1}{4\sqrt{|\ep|}} 
\sin 
\left( \fr {R}{\sqrt{|k_2+2k_3|}} 
\right)
\Bigg]$
&  &  \\ [20pt]
\hline \hline
\end{tabular}
\caption{Solutions for the Green function $\cG_1$
from equation \rf{cG1} in the (large amplitude) limit where $\ch=-1/4 +\ep$ for small $\ep$.
The sign choice for $\ep$ and the coefficient combination $k_2 + 2 k_3$ are listed in the right two columns.
Here, 
$\ps=|k_2+k_3|/|k_2+2 k_3|$
and the coefficient combinations 
$k_1$, $k_2$, and $k_3$
are defined in equation \rf{constants}.
Note that $\ps$ and $R/|k_2+2 k_3|$ must be of $O(1)$ or smaller for the 
approximation to be valid.
Error terms for these approximations are of order $\sqrt{\ep}$.}
\label{solntable2}
\end{table}

Several features are clear in this limit.
Firstly, 
as $\ep \rightarrow 0$, 
it can be shown that 3 of the 4 solutions in Table \ref{solntable2}
are finite. 
The fourth row, 
with ${\rm sgn} (k_2 +2 k_3 )=-1$ and ${\rm sgn} \ep =-1 $
diverges as $\ep \rightarrow 0$.
Second, 
the solutions are
oscillatory in $R$ with no damping 
(second row $-+$ case), 
a mixture of damped 
and oscillatory behavior in $R$ 
(third and fourth row $+-$ and $--$ cases), 
or damped with no oscillations (first row $++$ case).
The nature of the solution
depends critically on which part of the
$k_1$, 
$k_2$, 
$k_3$ coefficient space one probes.

\subsection{Special cases: $k_2^2-k_1 k_3=0 \vee k_1+2k_2=0$}
\label{Special cases}

We record here the solution for the Green function for the Newtonian potential when the coefficient combinations
$k_1$, $k_2$, and $k_3$ (\rf{constants}) 
take on special values.
When $k_2^2-k_1 k_3=0$, 
we cannot apply the solution \rf{cG1} directly.
We go back to re-evaluate the Fourier transform integral \rf{G1} with the $p^4$ term absent.
The result for $G_1$, and $G_2$,
the Green functions for ${\bar h}_{00}$
and ${\bar h}_{jj}$, 
are given by
\bea
G_1 (\vec r, \vec r^\prime ) &=& 
\fr {1}{\pi R} \left( 1 - \fr {k_1 + k_2}{k_1 + 2 k_2} 
e^{-i \frac{R}{\sqrt{k_1 + 2 k_2}} } \right),
\nonumber\\
G_2 (\vec r, \vec r^\prime ) &=& 
-\fr {1}{\pi R} \fr {k_2 + k_3}{k_1 + 2 k_2} 
e^{-i \frac{R}{\sqrt{k_1 + 2 k_2}} }.
\label{G1G2}
\eea
Note that when the sign of $k_1 + 2 k_2$
is negative, 
the solution becomes a damped exponential
of the Yukawa form.
The Green function for $h_{00}$ can be obtained
from $\cG_1=(1/2)(G_1 + G_2)$.

Also we consider here the special case
when $k_2^2-k_1 k_3=0$ and $k_1+2k_2=0$.
In this case the nonstandard terms in 
the momentum space functions in \rf{GreensolnMom} 
are constants, 
yielding delta functions with the Fourier transform.
The Green function $\cG_1$ is then given by
\beq
\cG_1 = \fr {1}{2\pi R} + 2 k_3 \de^{(3)} (\vec R).
\label{cGdelta}
\eeq


\section{References}

\bibliographystyle{unsrt}
\bibliography{refs.bib}

\end{document}